\documentclass[journal]{IEEEtran}
\usepackage{graphicx,amssymb,fancyhdr}
\usepackage{amsmath,amsfonts,amssymb}

\usepackage[usenames]{color}
\usepackage{bm,ulem,cite,color,epstopdf,mathrsfs}
\begin{document}
\bibliographystyle{IEEEtran}
\title{\LARGE{Soft Pilot Reuse and Multi-Cell Block Diagonalization Precoding for Massive MIMO Systems}}

\author{\IEEEauthorblockN{Xudong Zhu, Zhaocheng Wang, \textit{Senior Member, IEEE},
 Chen Qian, Linglong Dai, \textit{Senior Member, IEEE}, \\
 Jinhui Chen, Sheng Chen, \textit{Fellow, IEEE}, and Lajos Hanzo, \textit{Fellow, IEEE}} %

\thanks{Copyright \textcopyright 2015 IEEE. Personal use of this material is permitted.
 However, permission to use this material for any other purposes must be obtained from the IEEE
 by sending a request to pubs-permissions@ieee.org.} %

\thanks{The authors would like to thank the reviewers
 for their expert comments that helped to improve the paper.}%

\thanks{X. Zhu, Z. Wang, C. Qian, and L. Dai are with Tsinghua National Laboratory for
 Information Science and Technology (TNList), Department of Electronic Engineering,
 Tsinghua University, Beijing 100084, China (E-mails: zhuxd12@mails.tsinghua.edu.cn,
 zcwang@tsinghua.edu.cn, qianc10@mails.tsinghua.edu.cn, daill@tsinghua.edu.cn).}%

\thanks{J. Chen is with SONY China Research Laboratory, Beijing 100190, P. R. China
 (E-mail: jinhui.chen@sony.com.cn).}%

\thanks{S. Chen and L. Hanzo are with Electronics and Computer Science, University
 of Southampton, Southampton SO17 1BJ, U.K. (E-mails: sqc@ecs.soton.ac.uk, lh@ecs.soton.ac.uk).
 S. Chen is also with King Abdulaziz University, Jeddah 21589, Saudi Arabia.}%

\thanks{This work was supported by Beijing Natural Science Foundation (Grant No. 4142027),
 National High Technology Research and Development Program of China (Grant No. 2014AA01A704),
 National Nature Science Foundation of China (Grant No. 61271266),
 the Foundation of Shenzhen government,
 and Information Technology Development Division, Sony Corporation and Sony China Research Laboratory, Sony (China) Ltd.}
\vspace*{-10mm}
}
\maketitle

\begin{abstract}
 The users at cell edge of a massive multiple-input multiple-output
 (MIMO) system suffer from severe pilot contamination, which leads to
 poor quality of service (QoS). In order to enhance the QoS for
 these edge users, soft pilot reuse (SPR) combined with multi-cell
 block diagonalization (MBD) precoding are proposed.  Specifically,
 the users are divided into two groups according to their large-scale
 fading coefficients, referred to as the center users, who only suffer
 from modest pilot contamination and the edge users, who suffer from
 severe pilot contamination. Based on this distinction, the SPR scheme
 is proposed for improving the QoS for the edge users, whereby a
 cell-center pilot group is reused for all cell-center users in all
 cells, while a cell-edge pilot group is applied for the edge users in
 the adjacent cells. By extending the classical block
 diagonalization precoding to a multi-cell scenario, the MBD
 precoding scheme projects the downlink transmit signal onto the
 null space of the subspace spanned by the inter-cell channels of
 the edge users in adjacent cells. Thus, the inter-cell
 interference contaminating the edge users' signals in the adjacent
 cells can be efficiently mitigated and hence the QoS of these edge
 users can be further enhanced. Our theoretical analysis and
 simulation results demonstrate that both the uplink and downlink
 rates of the edge users are significantly improved, albeit at the
 cost of the slightly decreased rate of center users.
\end{abstract}

\IEEEpeerreviewmaketitle

\section{Introduction}\label{S1}

 In an effort to meet the escalating demand for increasingly
 higher-capacity and improved-reliability wireless systems, the
 `massive' or large-scale multiple-input multiple-output (LS-MIMO)
 concept has been proposed \cite{noncooperative}-\cite{LuLu}, where
 typically each base station (BS) is equipped with a large number of
 antenna elements (AEs) to serve a much smaller number of single-AE
 users. This way each user may have access to several AEs. This
 large-scale MIMO technology offers several significant advantages in
 comparison to the conventional MIMO concept having a moderate number
 of AEs. Firstly, asymptotic analysis based on random matrix theory
 \cite{scalingupMIMO} demonstrates that both the intra-cell
 interference and the uncorrelated noise effects can be efficiently
 mitigated, as the number of AEs tends to infinity.  Furthermore, the
 energy consumption of cellular BSs can be substantially reduced
 \cite{milliwatts} and the LS-MIMO systems are robust, since the
 failure of one or a few of the AEs and radio frequency (RF) chains
 would not appreciably affect the resultant system performance
 \cite{noncooperative}. Additionally, low-complexity signal-processing
 relying on matched-filter (MF) based transmit precoding (TPC) and
 detection can be used to for approaching the optimal performance,
 when the number of AEs at the BS tends to
 infinity~\cite{scalingupMIMO}.

 Similar to conventional MIMO systems, knowledge of the channel state
 information (CSI) is also required at the BS of LS-MIMO systems,
 namely for data detection in the uplink (UL) and for multi-user TPC in
 the downlink (DL)~\cite{scalingupMIMO,conventionalMIMO}.
 In the time-division duplexing (TDD) protocol, the BS estimates the UL
 channels and obtains the DL CSI by exploiting the channel's
 reciprocity~\cite{noncooperative,LuLu,TDD1}.  However, this approach
 suffers from the so-called pilot contamination (PC) problem
 \cite{noncooperative,scalingupMIMO,LuLu} in multi-cell multi-user
 scenarios due to the reuse of the pilot sequences in adjacent cells,
 which imposes grave interference on the channel estimate at the
 BS. Furthermore, the commonly used MF and zero-forcing (ZF) TPC
 schemes will impose inter-cell interference (ICI) on the DL
 transmission, which cannot be reduced by increasing the number of AEs
 at the BS.

 Hence, the problem of ICI and PC has been extensively studied
 \cite{FFR,2G5G,ComP1,ComP2,timeshift1,timeshift2,PCP1,PCP2,channelcovariance,AOA1,AOA2,dataaided,blind1,blind2,Zhang_etal2014}.
 The fractional frequency reuse (FFR) scheme~\cite{FFR,2G5G}
 adopted in LTE Release 9 aims for mitigating the ICI by
 assigning orthogonal frequency bands to edge users in the adjacent
 cells at the cost of additional spectral resources.
 The original frequency-division duplexing (FDD) based coordinated
 multi-point (CoMP) transmission of LTE-A Release 11
 \cite{ComP1} is able to avoid the ICI between adjacent cells,
 whereby each user estimates and feeds back the quantized DL channel
 from all adjacent cells to the corresponding BS, and then the BS distributes the CSI
 to adjacent cells. However, this kind of FDD based CoMP technique
 would not be feasible for massive MIMO since the CSI feedback overhead
 would be huge as the number of BS antennas increasing \cite{ComP2}.
 Using time-shifted pilot sequences for asynchronous transmission
 among the adjacent cells~\cite{timeshift1,timeshift2} partially
 mitigates this problem, but it leads to mutual interference between
 data transmission and pilot transmission. A TPC scheme can be used
 for mitigating the ICI with the aid of joint multi-cell
 processing~\cite{PCP1,PCP2}, but again, imposes a high information
 exchange overhead.  The authors of~\cite{channelcovariance} imposed
 specific conditions on the channel's covariance matrix, which is only
 valid for the asymptotic case of infinitely many AEs at the BS. The
 angle-of-arrival (AOA) based methods of~\cite{AOA1,AOA2} exploit the
 fact that the users having mutually non-overlapping AOAs hardly
 contaminate each other even if they use the same pilot sequence, but
 naturally, the efficiency of these methods relies on the assumption
 that the AOA spread of each user is small, which is not always the
 case under realistic channel conditions. A data-aided channel
 estimation scheme was proposed in~\cite{dataaided}, whereby partially
 decoded data is used for estimating the channel and the PC effects
 can be beneficially reduced by iterative processing at the cost of an
 increased computational complexity. Additionally, the blind method
 of~\cite{blind1,blind2} based on subspace partitioning is capable of
 reducing the ICI under the assumption that the channel vectors of
 different users are orthogonal, which is not often the case in
 practice. The scheme proposed in \cite{Zhang_etal2014} is capable of
 eliminating PC all together, but this is achieved with the aid of a
 complex DL and UL training procedure. Note that all these existing
 contributions treat all users in the same way, as though they suffer
 from the same PC, but in reality the severity of PC varies among the
 users.

 Against the above background, inspired by the FFR
 scheme~\cite{FFR} adopted in LTE Release 9, we propose a
 soft pilot reuse (SPR) scheme for mitigating the PC of LS-MIMO systems,
 whereby a cell-edge pilot group is applied for the cell-edge users in
 adjacent cells, while the cell-center users reuse the same center
 pilot group in all cells. Furthermore, by extending the
 classical block diagonalization (BD) precoding \cite{BD} to
 a multi-cell scenario, a multi-cell block diagonalization (MBD)
 TPC techniques is conceived for mitigating the ICI and for enhancing the
 quality of service (QoS) for the edge users. Specifically, the
 contributions of this paper are summarized as follows.

\begin{itemize}
\item We break away from the traditional practice of treating the PC
  for all users identically - instead, we divide the users into two
  different groups to be considered separately, namely, center users
  subjected to a slight PC and the edge users suffering from more
  severe PC. In this way, the center users can benefit directly from
  the LS-MIMO technology and the efforts can be directed towards
  improving the QoS for the edge users.
\item In contrast to the FFR scheme, which assigns orthogonal
  frequency bands to the edge users in adjacent cells, the proposed SPR
  scheme divides the pilot types into two groups within the same
  frequency band, i.e. in a center pilot group, which is reused for the
  center users in all cells and in an edge pilot group, which is applied
  for the edge users in adjacent cells. Thus, for the edge users, the
  accuracy of the channel estimation is improved and the UL achievable
  rate is increased. Moreover, by using slightly more pilot resources
  for edge users, the BS becomes capable of estimating not only the
  intra-cell channels of the users within the reference cell, but also
  the knowledge of the `inter-cell channels' of the edge users in the
  adjacent cells.
\item Different from the original CoMP technique has
  to obtain the inter-cell channels by consuming large overhead \cite{ComP1,ComP2},
  the proposed MBD precoding can directly exploit the partial
  knowledge of the `inter-cell channels' and is capable of suppressing
  the ICI imposed on the edge users of the adjacent
  cells. Specifically, by extending the classical BD TPC to a
  multi-cell scenario, the MBD TPC projects the DL transmit signal
  onto the null space of the subspace spanned by the partially known
  `inter-cell channels'. Thus, the ICI imposed on the edge users of
  the adjacent cells can be substantially mitigated, hence the QoS of
  the edge users is significantly enhanced.
\item In order to analyze the performance of our proposal, we compare
  the associated pilot requirements, derive the attainable average UL
  as well as DL rate and characterize the computational complexity
  imposed. Our theoretical derivation confirms that both the
  achievable UL and DL rate of the edge users is significantly
  improved at the cost of requiring slightly more
  pilots. Moreover, our simulation results show that the
  average UL and DL cell throughput in the SPR and MBD aided system
  is able to approach and even exceed that of the conventional
  system, provided that a modestly increased number of BS AEs is affordable.
\end{itemize}

 The rest of the paper is organized as follows. In Section~\ref{S2},
 we briefly review the multi-cell LS-MIMO system model, while
 Section~\ref{S3} is devoted to detailing the PC, which is the main
 performance-limiting factor of LS-MIMO systems. Section~\ref{S4}
 further details the motivation of this paper, while the proposed SPR scheme
 and MBD precoding are discussed in Section~\ref{S5}.  Section~\ref{S6}
 provides our performance analysis of the proposed SPR scheme and MBD
 precoding. Our simulation results quantifying the benefits of our proposals
 are presented in Section~\ref{S7}, while our conclusions follow in Section~\ref{S7}.

 Throughout our discussions, boldface lower and upper-case symbols
 represent vectors and matrices, respectively. The transpose,
 conjugate, and Hermitian transpose operators are given by $(\cdot
 )^T$, $(\cdot )^*$, and $(\cdot )^H$, respectively. The Moore-Penrose
 pseudo inverse operator is denoted by $(\cdot )^{\dagger}$ and the
 trace operator is represented by $\text{Tr}(\cdot)$, while
 $\mbox{diag}\{a_1,a_2,\cdots , a_m\}$ denotes the diagonal matrix
 associated with $a_1,a_2,\cdots , a_m$ at its diagonal entries and
 the $M\times M$ identity matrix is given by $\mathbf{I}_M$. The
 number of elements in a set is denoted by $\text{card}\{\cdot\}$, and
 the $l_p$ norm is denoted by $\|\cdot \|_p$, while the expectation
 operator is given by $\text{E}\{\cdot \}$.

\begin{figure}
\center{\includegraphics[angle=90,width=0.45\textwidth]{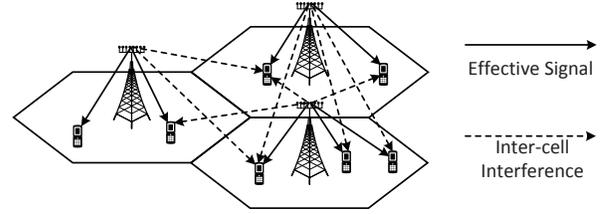}}
\vspace*{-2mm}
\caption{Illustration of multi-user multi-cell LS-MIMO system.}
\label{systemmodel} 
\vspace*{-4mm}
\end{figure}

\section{System Model}\label{S2}

 A multi-cell multi-user LS-MIMO system is illustrated in
 Fig.~\ref{systemmodel}, which is composed of $L$ hexagonal cells,
 each having a central BS associated with $M$ antennas to serve $K$
 ($K\ll M$) single-antenna users \cite{noncooperative,scalingupMIMO}.
 The channel vector $\mathbf{h}_{i,j,k}\in\mathbb{C}^{M\times 1}$
 of the link spanning from the $k$-th user of the $j$-th cell to
 the BS of the $i$-th cell can be formulated as
\begin{equation}\label{eq1} 
 \mathbf{h}_{i,j,k} = \mathbf{g}_{i,j,k}\sqrt{\beta_{i,j,k}} .
\end{equation}
 The small-scale fading vectors
 $\mathbf{g}_{i,j,k}\in\mathbb{C}^{M\times 1}$ are statistically
 independent for the $K$ users and they obey the complex-valued
 Gaussian distribution having a zero-mean vector and a covariance
 matrix $\mathbf{I}_M$, hence we have
 $\mathbf{g}_{i,j,k}\sim\mathcal{CN}\big(\mathbf{0},\mathbf{I}_M\big)$.
 Still referring to Eq.~\ref{eq1}, the large-scale fading coefficients
 $\beta_{i,j,k}$ are the same for the different antennas at the same
 BS, but they are user-dependent. Moreover, they are related to both the
 pathloss and shadow fading, which will be addressed in detail in the context of
 our simulations. Thus, the channel matrix of all the $K$ users in the
 $j$-th cell and the BS in the $i$-th cell can be represented by
\begin{align}\label{channeldefine} 
 \mathbf{H}_{i,j}=& \big[\mathbf{h}_{i,j,1} ~ \mathbf{h}_{i,j,2} \cdots \mathbf{h}_{i,j,K}\big]\nonumber\\
 =& \big[\mathbf{g}_{i,j,1} ~ \mathbf{g}_{i,j,2} \cdots \mathbf{g}_{i,j,K}\big]
 \mathbf{D}_{i,j}^{1/2},
\end{align}
 where $\mathbf{D}_{i,j}=\text{diag}(\beta_{i,j,1},\beta_{i,j,2},\cdots ,\beta_{i,j,K})$
 denotes the large-scale fading matrix relating all the  $K$ users in the  $j$-th cell to
 the $i$-th cell's BS.

 By adopting the TDD protocol, the BS obtains the DL channel estimate
 by exploiting the reciprocity of the UL and DL channels. More specifically,
 both the small-scale fading vectors and the large-scale fading coefficients
 may be deemed to be equal for both the DL and UL directions, provided
 that the bandwidth is sufficiently narrow for avoiding the
 independent fading of the DL and UL.

 Before considering the PC phenomenon, we summarize the asymptotic
 orthogonality on random vector \cite{noncooperative}. Let
 $\mathbf{x},\mathbf{y}\in\mathbb{C}^{M\times 1}$ be two independent
 vectors with distribution
 $\mathcal{CN}(\mathbf{0},c\mathbf{I}_M)$. Then from the law of large
 numbers, we have
\begin{equation} 
 \lim_{M\rightarrow\infty}\frac{\mathbf{x}^H\mathbf{x}}{M}\overset{a.s.}{\longrightarrow}c
 \hspace{0.4cm}\text{and}\hspace{0.4cm}
 \lim_{M\rightarrow\infty}\frac{\mathbf{x}^H\mathbf{y}}{M}\overset{a.s.}{\longrightarrow}0,
\end{equation}
 where $\overset{a.s.}{\longrightarrow}$ denotes the almost sure convergence.

\section{Pilot Contamination}\label{S3}

\begin{figure}
\center{\includegraphics[angle=90,width=0.4\textwidth]{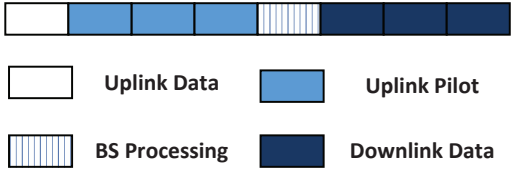}}
\vspace*{-2mm}
\caption{Multi-user multi-cell MIMO TDD protocol.}
\label{TDD} 
\vspace*{-3mm}
\end{figure}

 By considering the TDD protocol, we adopt the widely used block-fading channel model,
 whereby the channel vectors $\mathbf{h}_{i,j,k}$ remain constant during the channel's
 coherence interval. As shown in Fig.~\ref{TDD}, each coherence interval is
 comprised of four stages for each user \cite{timeshift2}: 1)~UL data
 transmission; 2)~UL pilot transmission; 3)~BS processing; and 4)~DL data
 transmission.

 At the first stage, all users in all cells synchronously send UL data to their
 corresponding BSs and the user data received at the BS in the $i$-th cell can be
 represented as
 $\mathbf{y}_i^{\text{u}} = \sqrt{\rho_\text{u}}\sum_{j=1}^{L}\sum_{k=1}^{K}
 \mathbf{h}_{i,j,k}x_{j,k}^{\text{u}} + \mathbf{n}_i^{\text{u}}$,
 where $x_{j,k}^{\text{u}}$ with
 $\text{E}\{|x_{j,k}^{\text{u}}|^2\}=1$ denotes the symbol transmitted
 from the $k$-th user roaming in the $j$-th cell, $\rho_{\text{u}}$
 represents the UL data transmission power and
 $\mathbf{n}_i^{\text{u}}\in\mathbb{C}^{M\times 1}$ denotes the
 corresponding UL channel's additive Gaussian white noise (AWGN) vector
 associated with
 $E\big\{\mathbf{n}_i^{\text{u}}\big(\mathbf{n}_i^{\text{u}}\big)^H\big\}=
 \big(\sigma_n^{\text{u}}\big)^2\mathbf{I}_M$.

 For a typical LS-MIMO system, the pilot sequences used
 within a specific cell are orthogonal, but the same pilot group is
 typically reused in the adjacent cells due to the limited number of
 orthogonal pilot sequences. Thus, during the second stage, the matrix
 of pilot sequences received at the BS of the $i$-th cell,
 which is denoted by $\mathbf{Y}_i^{\text{p}}
 \in\mathbb{C}^{M\times \tau}$, can be represented as
 $\mathbf{Y}_i^{\text{p}} = \sqrt{\rho_{\text{p}}}\sum_{j=1}^{L}\mathbf{H}_{i,j}
 \mathbf{\Phi}+\mathbf{N}_i^{\text{p}}$,
 where the matrix $\mathbf{\Phi}=\big [\bm{\phi}_1 ~
   \boldsymbol{\phi}_2 \cdots
   \boldsymbol{\phi}_K\big]^T\in\mathbb{C}^{K\times\tau}$ containing
 the transmitted pilot sequence satisfies
 $\mathbf{\Phi}\mathbf{\Phi}^H=\mathbf{I}_K$, $\rho_{\text{p}}$ is the
 transmission power of the pilots and
 $\mathbf{N}_i^{\text{p}}\in\mathbb{C}^{M\times\tau}$ denotes the UL
 channel's AWGN matrix.

 During the third stage, the BS of the $i$-th cell obtains an estimate of the channel matrix
 $\mathbf{H}_{i,i}$ using any conventional channel estimation method by directly
 correlating the received pilot matrix with the local pilot matrix, yielding
\begin{equation}\label{eq6} 
 \widehat{\mathbf{H}}_{i,i}=\frac{1}{\sqrt{\rho_{\text{p}}}}\mathbf{Y}_i^{\text{p}}\mathbf{\Phi}^H
 =\mathbf{H}_{i,i}+\sum_{j\neq i}\mathbf{H}_{i,j} +
 \frac{1}{\sqrt{\rho_{\text{p}}}}\mathbf{N}_i^{\text{p}}\mathbf{\Phi}^H.
\end{equation}
 It can readily be seen that the channel estimate of the $k$-th user
 in the $i$-th cell, namely $\widehat{\mathbf{h}}_{i,i,k}$, is a
 linear combination of the channels $\mathbf{h}_{i,j,k}$ for $1\le
 j\le L$, which include the channels of the users in the other cells
 associated with the same pilot sequence. This phenomenon is referred
 to as PC~\cite{noncooperative,scalingupMIMO,LuLu}. Given the
 estimated channel matrix $\widehat{\mathbf{H}}_{i,i}$ and by adopting
 the low-complexity MF detector, the detected symbol arriving from the
 $k$-th user in the $i$-th cell can be represented as
\begin{align}\label{ULdatadetection} 
 \widehat{x}_{i,k}^{\text{u}} =& \widehat{\mathbf{h}}_{i,i,k}^H\mathbf{y}_i^{\text{u}} \nonumber\\
 =& \sqrt{\rho_{\text{u}}}\left(\mathbf{h}_{i,i,k}^H\mathbf{h}_{i,i,k} x_{i,k}^{\text{u}}
 + \sum_{j\neq i}\mathbf{h}_{i,j,k}^H\mathbf{h}_{i,j,k} x_{j,k}^{\text{u}}\right)
 + \varepsilon_{i,k}^{\text{u}} \nonumber\\
 \overset{(a)}\approx & M\sqrt{\rho_{\text{u}}}\left(\beta_{i,i,k} x_{i,k}^{\text{u}}
 + \sum_{j\neq i}\beta_{i,j,k} x_{j,k}^{\text{u}}\right),
\end{align}
 where $\mathbf{v}_{i,k}$ denotes the $k$-th column of
 $\frac{1}{\sqrt{\rho_{\text{p}}}}
 \mathbf{N}_i^{\text{p}}\mathbf{\Phi}^H$,
 $\varepsilon_{i,k}^{\text{u}}$ represents the interference, which can
 be reduced to an arbitrarily low level by increasing the number of
 transmit antennas $M$ at the BS, and $\overset{(a)}\approx$ indicates
 that the approximation holds by invoking the asymptotic orthogonality
 associated with $M\rightarrow\infty$. Thus, the UL signal to
 interference plus noise ratio (SINR) of the $k$-th user in the $i$-th
 cell can be calculated as
\begin{align}\label{SINRUL} 
 \text{SINR}_{i,k}^{\text{u}} =& \frac{|\mathbf{h}_{i,i,k}^H\mathbf{h}_{i,i,k}|^2}
 {\sum_{j\neq i}|\mathbf{h}_{i,j,k}^H\mathbf{h}_{i,j,k}|^2 +|\varepsilon_{i,k}^{\text{u}}|^2/\rho_{\text{u}}}\nonumber\\
 \overset{(a)}\approx & \frac{\beta_{i,i,k}^2}{\sum_{j\neq i}\beta_{i,j,k}^2} ,
\end{align}
 and the achievable UL rate can be expressed as
 $\text{C}_{i,k}^{\text{u}}
 =(1-\mu)\text{E}\big\{\log_2\big(1+\text{SINR}_{i,k}^{\text{u}}\big)\big\}$, where $0<\mu<1$ evaluates the spectral efficiency reduction caused by the pilot transmission \cite{Overhead}.
 It is clear the UL achievable rate remains limited by
 the PC and it cannot be increased by simply assigning an increased
 transmission power and/or pilot power, i.e., by increasing
 $\rho_{\text{u}}$ and/or $\rho_{\text{p}}$.

 The PC affects the DL transmission during the fourth stage as well.
 The normalized MF precoding matrix~\cite{LuLu} is commonly used for the DL
 transmission, which can be represented by
 $\mathbf{W}_i=\frac{1}{\sqrt{\gamma_i}} \widehat{\mathbf{H}}_{i,i}^*$,
 where $\gamma_i=\text{Tr}\big(\widehat{\mathbf{H}}_{i,i}^T\widehat{\mathbf{H}}_{i,i}^*\big)/K$
 is a normalization factor. The BS in the $i$-th cell transmits an $M$-dimensional
 signal vector as $\mathbf{s}_i^{\text{d}}=\mathbf{W}_i\mathbf{x}_{i}^{\text{d}}$, where
 $\mathbf{x}_{i}^{\text{d}}=\big[x_{i,1}^{\text{d}} ~ x_{i,2}^{\text{d}} \cdots
 x_{i,K}^{\text{d}}\big]^T$ with $\text{E}\big\{|x_{i,k}^{\text{d}}\big|^2\}=1$ denotes
 the source symbol vector for the $K$ users in the $i$-th cell. The received signals of
 the $K$ users in the $i$-th cell can be collected together as
 $\mathbf{y}_{i}^{\text{d}} =
 \sqrt{\rho_{\text{d}}}\sum_{j=1}^L \mathbf{H}_{j,i}^T \frac{1}{\sqrt{\gamma_j}}
 \widehat{\mathbf{H}}_{j,j}^*\mathbf{x}_j^{\text{d}}+\mathbf{n}_i^{\text{d}}$,
 where $\mathbf{n}_i^{\text{d}}$ denotes the DL channel AWGN vector associated with
 $E\big\{\mathbf{n}_i^{\text{d}}\big(\mathbf{n}_i^{\text{d}}\big)^H\big\}=
 \big(\sigma_n^{\text{d}}\big)^2\mathbf{I}_M$. Similar to the
 derivation seen in (\ref{SINRUL}), the DL SINR of the $k$-th user in the $i$-th cell
 can be derived as
\begin{align}\label{SINRDL} 
 \text{SINR}_{i,k}^{\text{d}} =& \frac{|\mathbf{h}_{i,i,k}^T\mathbf{h}_{i,i,k}^{*}|^2}
 {\sum_{j\neq i}|\mathbf{h}_{j,i,k}^T\mathbf{h}_{j,i,k}^{*}|^2 +|\varepsilon_{i,k}^{\text{d}}|^2/\rho_{\text{d}}}\nonumber\\
 \overset{(a)}\approx & \frac{\beta_{i,i,k}^2}{\sum_{j\neq i}\beta_{j,i,k}^2},
\end{align}
 where $\varepsilon_{i,k}^{\text{d}}$ denotes the corresponding interference similar to
 $\varepsilon_{i,k}^{\text{u}}$ given in (\ref{ULdatadetection}). The corresponding
 DL rate can be represented as $\text{C}_{i,k}^{\text{d}}=(1-\mu)\text{E}\big\{\log_2
 \big(1+\text{SINR}_{i,k}^{\text{d}}\big)\big\}$.

 In summary, the PC caused by the reuse of the same orthogonal pilot
 group in adjacent cells cannot be reduced by increasing the number of
 antennas at the BS, hence it limits the achievable performance of
 multi-cell multi-user LS-MIMO systems.

\section{Motivation of Our Proposal}\label{S4}

 In the existing state-of-the-art solutions
 \cite{FFR,2G5G,timeshift1,timeshift2,PCP1,PCP2,channelcovariance,AOA1,AOA2,dataaided,blind1,blind2,Zhang_etal2014},
 which aim for reducing the PC, all users are treated
 identically. However, according to (\ref{SINRUL}) and (\ref{SINRDL}),
 it becomes clear that the attainable SINR is proportional to the
 large-scale fading coefficients $\beta_{i,i,k}^2$, which are
 different for the $K$ users of each cell. Thus, we have to break
 away from this traditional concept of treating the PC for all users
 identically, which this motivates our idea of dividing the users of each
 cell into two groups, namely the group of center users subjected to modest
 PC and the group of edge users suffering from severe PC. We will treat them
 differently.

 In fact, the limit of the UL SINR of the $k$-th user in the $i$-th cell, which is
 defined by
\begin{equation}\label{lambda} 
 \eta_{i,k}=\frac{\beta_{i,i,k}^2}{\sum_{j\neq i}\beta_{i,j,k}^2} ,
\end{equation}
 specifies the severity of the PC for this user. Therefore, it is easy
 to sort the users in a cell according to their SINR values
 $\eta_{i,k}$, if all the large-scale fading coefficients
 $\{\beta_{i,j,k}^2\}$ are known at the BS, which is a key assumption
 stipulated in the state-of-the-art
 contributions~\cite{TDD1,timeshift1,dataaided}. However, in practice,
 it is difficult for the BS to obtain an accurate estimate of the
 large-scale fading coefficients of the users in other cells, i.e., of
 $\beta_{i,j,k}^2$ for $j\neq i$, unless BS-cooperation is invoked,
 which is typically associated with a substantial side-information
 overhead.

\begin{figure}
\center{\includegraphics[angle=0,width=0.45\textwidth]{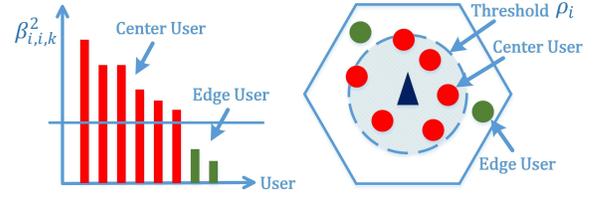}}
\vspace*{-2mm}
\caption{An illustrative example of the user division.}
\label{CellDivision} 
\vspace*{-3mm}
\end{figure}

 Since we have $\eta_{i,k} \propto \beta_{i,i,k}^2$, we may also use
 $\beta_{i,i,k}^2$ for estimating the severity of the PC for the
 $k$-th user roaming in the $i$-th cell.  In contrast to
 $\beta_{i,j,k}^2$ for $j\neq i$, all the large-scale fading
 coefficients $\{\beta_{i,i,k}^2\}$ of the $K$ users in the $i$-th
 cell can be readily obtained. Thus, the $K$ users in the $i$-th cell
 can be readily divided into two groups according to
\begin{equation}\label{eq13} 
 \beta_{i,i,k}^2\overset{?}>\rho_i\rightarrow
  \left\{ \begin{array}{ll}
   \text{Yes}\rightarrow \text{center users}, \\
   \text{No}\hspace{0.1cm}\rightarrow \text{edge users}.
  \end{array}
 \right.
\end{equation}
 The user-grouping threshold $\rho_i$ can be set to
\begin{equation}\label{eq14} 
 \rho_i=\frac{\lambda}{K}\sum_{k=1}^K\beta_{i,i,k}^2,
\end{equation}
 where $\lambda$ can be adjusted according to the specific system
 configuration.  A simple case is illustrated in
 Fig.~\ref{CellDivision}, where according to the large-scale fading
 coefficients $\{\beta_{i,i,k}^2\}$ and the given threshold $\rho_i$,
 the users are divided into two groups, namely the center users
 associated with only a slight PC and the edge users subjected to
 severe PC. Note that the threshold $\rho_i$ is not based on the
 geographic locations of the users - it is rather based on the signal
 space of $\{\beta_{i,i,k}^2\}$.

 Since the center users only suffer from minor PC, the conventional
 LS-MIMO scheme outlined in the previous section is capable of
 attaining a high performance. By contrast, the edge users suffer from
 serious pilot contamination, hence their performance based on the
 conventional LS-MIMO scheme is expected to be poor. In order to
 enhance the QoS of the edge users, who suffer from heavy PC, we
 propose the more sophisticated SPR scheme and MBD precoding in the following
 section.

\section{The Proposed Soft Pilot-Reuse Scheme and
Multi-cell Block Diagonalization Precoding}\label{S5}

 Based on the division of users into two groups as outlined in the
 previous section, it is plausible that the center users indeed
 benefit from the conventional LS-MIMO technique. By contrast,
 improved measures have to be considered for enhancing the QoS of the
 edge users, such as our SPR and MBD schemes, which will be discussed
 in detail in the following three subsections: 1)~the proposed SPR scheme;
 2)~channel estimation based on SPR; and 3)~the MBD precoding advocated.

\subsection{Proposed Soft Pilot Reuse Scheme}\label{S5.1}

   Inspired by the FFR scheme, which assigns orthogonal frequency
   bands to edge users in adjacent cells to prevent serious ICI in
   3GPP LTE Release 9, we propose the SRP scheme to mitigate the PC,
   whereby orthogonal pilot sub-groups are assigned to the edge
   users in the adjacent cells, while a center pilot group is reused
   for the center users of all cells.

 More specifically, consider a typical LS-MIMO system, which is
 composed of $L$ hexagonal cells, where the $i$-th cell supports $K_i$
 users. In the conventional LS-MIMO scheme
 \cite{noncooperative,scalingupMIMO,LuLu}, the number of orthogonal
 pilot sequences required can be calculated as
\begin{equation}\label{KCS} 
 K_{\text{CS}} =\max\{K_i,\hspace{0.2cm}i=1,2,\cdots,L\}.
\end{equation}
 In contrast to the conventional LS-MIMO scheme, where all users are
 treated identically, the $K_i$ users of the $i$-th cell are firstly
 divided into two groups according to their large-scale fading
 coefficients $\{\beta_{i,i,k}^2\}$, which have cardinalities of
\begin{equation} 
 K_i = K_{i,\text{c}} + K_{i,\text{e}},
\end{equation}
 where $K_{i,\text{c}}=\text{card}\{k: \beta_{i,i,k}^2>\rho_i\}$ denotes the number
 of center users, while $K_{i,\text{e}}=\text{card}\{k: \beta_{i,i,k}^2\leq\rho_i\}$
 represents the number of edge users. Thus, the number of orthogonal pilot sequences
 needed in the proposed SPR scheme can be calculated as
\begin{equation}\label{KSPR} 
 K_{\text{SPR}} = K_{\text{c}} + K_{\text{e}},
\end{equation}
 where $K_{\text{c}}=\max\{K_{i,\text{c}}, i=1,2,\cdots,L\}$ denotes
 the number of pilot sequences assigned to the center users, while
 $K_{\text{e}}=\sum_{i=1}^{L}K_{i,\text{e}}$ denotes the number of
 pilot sequences dedicated to the edge users.
 It should be pointed out that we assume having $L$ cooperating
 cells, thus $L$ is a moderate value. For example, we have $L$=7
 for the classic seven-cell system.
 Then the entire set of pilot
 resources
 $\mathbf{\Phi}_{\text{SPR}}\in\mathbb{C}^{K_{\text{SPR}}\times \tau}$
 associated with
 $\mathbf{\Phi}_{\text{SPR}}\mathbf{\Phi}_{\text{SPR}}^H=\mathbf{I}_{K_{\text{SPR}}}$
 can be divided into
\begin{equation} 
\mathbf{\Phi}_{\text{SPR}}=\left[\mathbf{\Phi}_{\text{c}}^T ~
 \mathbf{\Phi}_{\text{e}}^T\right]^T ,
\end{equation}
 where $\mathbf{\Phi}_{\text{c}}\in\mathbb{C}^{K_{\text{c}}\times\tau}$ is reused for
 the center users in all cells and $\mathbf{\Phi}_{\text{e}}\in\mathbb{C}^{K_{\text{e}}
 \times\tau}$ is applied to the edge users of the adjacent cells. Furthermore,
 $\mathbf{\Phi}_{\text{e}}$ can be divided into $L$ partitions, as
\begin{equation} 
 \mathbf{\Phi}_{\text{e}}=\left[\mathbf{\Phi}_{\text{e},1}^T ~ \mathbf{\Phi}_{\text{e},2}^T
 \cdots \mathbf{\Phi}_{\text{e},L}^T\right]^T ,
\end{equation}
 where
 $\mathbf{\Phi}_{\text{e},i}\in\mathbb{C}^{K_{i,\text{e}}\times\tau}$
 is applied to the $K_{i,\text{e}}$ edge users in the $i$-th
 cell. Thus, the pilot sequences applied to edge users are orthogonal
 to those of the other users roaming in the adjacent cells.

\begin{figure}
\vspace*{2mm}
\center{\includegraphics[angle=0,width=0.35\textwidth]{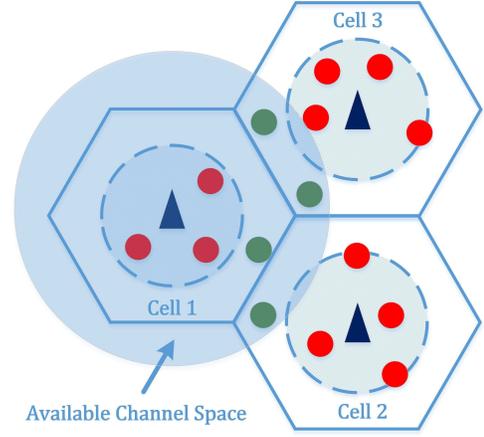}}
\caption{An example of the proposed SPR scheme.}
\label{InterCellChannel} 
\vspace*{-3mm}
\end{figure}

 In the example of Fig.~\ref{InterCellChannel}, there are 3 hexagonal
 cells associated with $K_1=4$, $K_2=5$, and $K_3=6$ users. In order
 to completely eliminate the PC would require 15 orthogonal pilot
 sequences. It can be readily calculated that we have
 $K_{\text{CS}}=6$ and $K_{\text{SPR}}=8$ for this simple
 case. Although the proposed SPR scheme requires slightly more pilot
 resources than the conventional scheme, the QoS of the edge users can
 be significantly improved, which will be verified in the following
 subsections.

\subsection{Channel Estimation Based on Soft Pilot Reuse}\label{S5.2}

 By applying the proposed SPR scheme, the BS becomes capable of
 estimating the channels for its edge users in the absence of PC,
 since the pilot sequences assigned to the edge users are all
 orthogonal. Moreover, the BS can also obtain the partial knowledge of
 the inter-cell channels of the edge users in adjacent cells, which is
 the dominant source of the ICI inflicted upon
 these edge users of the adjacent cells during the BS's DL
 transmissions.

 Specifically, we consider the same LS-MIMO system as in
 Subsection~\ref{S5.1}, which is composed of $L$ hexagonal cells, where
 the $i$-th cell has $K_i$ users. Based on the proposed SPR scheme, we
 divide the channel matrix $\mathbf{H}_{i,j}$ as defined in
 (\ref{channeldefine}) into two parts
\begin{equation}\label{eq20} 
 \mathbf{H}_{i,j}\rightarrow \big[\mathbf{H}_{i,j}^{\text{c}} ~ \mathbf{H}_{i,j}^{\text{e}}\big],
\end{equation}
 where $\mathbf{H}_{i,j}^{\text{c}}\in\mathbb{C}^{M\times
   K_{j,\text{c}}}$ denotes the channel matrix of the link spanning
 from the center users in the $j$-th cell to the BS in the $i$-th
 cell, while $\mathbf{H}_{i,j}^{\text{e}}\in\mathbb{C}^{M\times
   K_{j,\text{e}}}$ denotes the channel matrix of the link spanning
 from the edge users in the $j$-th cell to the BS in the $i$-th
 cell. Then, the pilot sequence received at the BS of the $i$-th cell can be represented by
\begin{align}\label{eq21}
 \overline{\mathbf{Y}}_i^{\text{p}} =& \sqrt{\rho_{\text{p}}}\left(\sum_{j=1}^{L}
 \mathbf{H}_{i,j}^{\text{c}}\mathbf{\Phi}_{\text{c}}\big(\text{r}:K_{j,\text{c}}\big) +
 \sum_{j=1}^{L}\mathbf{H}_{i,j}^{\text{e}}\mathbf{\Phi}_{\text{e},j}\right) + \overline{\mathbf{N}}_i^{\text{p}},
\end{align}
 where $\mathbf{\Phi}_{\text{c}}\big(\text{r}:K_{j,\text{c}}\big)$
 denotes the sub-matrix comprised of the first $K_{j,\text{c}}$ rows
 of $\mathbf{\Phi}_{\text{c}}$, while
 $\overline{\mathbf{N}}_i^{\text{p}}$ denotes the corresponding AWGN
 matrix at the UL receiver.

 Then the BS becomes capable of estimating the channel of its center users as
\begin{align}\label{eq22}
 \widehat{\mathbf{H}}_{i,i}^{\text{c}} =& \frac{1}{\sqrt{\rho_{\text{p}}}}
 \overline{\mathbf{Y}}_i^{\text{p}}\mathbf{\Phi}_{\text{c}}^H\big(\text{r}:K_{i,\text{c}}\big)\nonumber\\
 =& \mathbf{H}_{i,i}^{\text{c}} + \sum_{j\neq i}\mathbf{H}_{i,j}^{\text{c}}\big(\text{c}:K_{i,\text{c}}\big)
 + \overline{\mathbf{N}}_i^{\text{c}},
\end{align}
 where
 $\overline{\mathbf{N}}_i^{\text{c}}=\frac{1}{\sqrt{\rho_{\text{p}}}}
 \overline{\mathbf{N}}_i^{\text{p}}\mathbf{\Phi}_{\text{c}}^H\big(\text{r}:K_{i,\text{c}}\big)$
 which can be reduced to an arbitrarily small value by increasing $M$,
 and $\mathbf{H}_{i,j}^{\text{c}}\big(\text{c}:K_{i,\text{c}}\big)$
 denotes the matrix comprised of the first $K_{i,\text{c}}$ columns of
 $\mathbf{H}_{i,j}^{\text{c}}$.  Note that if we have $K_{i,\text{c}}
 > K_{j,\text{c}}$, then $K_{i,\text{c}} - K_{j,\text{c}}$ zero
 vectors are used to fill
 $\mathbf{H}_{i,j}^{\text{c}}\big(\text{c}:K_{i,\text{c}}\big)$.  In
 contrast to the conventional scheme of (\ref{eq6}), the proposed
 scheme only allows the center users to reuse the same pilot group of
 $\mathbf{\Phi}_{\text{c}}$, since the PC imposed on these center
 users is modest. Consequently, the severity of the PC inflicted upon
 the channel estimation of the center users given in (\ref{eq22}) is
 minor.

 On the other hand, by adopting the proposed SPR scheme, the BS of the $i$-th cell becomes
 capable of acquiring the channel estimate of its edge users without excessive PC,
 yielding
\begin{equation}\label{eq23}
 \widehat{\mathbf{H}}_{i,i}^{\text{e}}=\frac{1}{\sqrt{\rho_{\text{p}}}}
 \overline{\mathbf{Y}}_i^{\text{p}}\mathbf{\Phi}_{\text{e},i}^H
 =\mathbf{H}_{i,i}^{\text{e}} + \overline{\mathbf{N}}_i^{\text{e}},
\end{equation}
 where we have $\overline{\mathbf{N}}_i^{\text{e}}=\frac{1}{\sqrt{\rho_{\text{p}}}}
 \overline{\mathbf{N}}_i^{\text{p}}\mathbf{\Phi}_{\text{e},i}^H$, which can be made
 arbitrarily small by increasing the number of antennas at the BS. It is clear that
 the PC is completely eliminated for these edge users and, therefore,
 the channel estimation accuracy of these edge users is significantly enhanced. By
 contrast, with the conventional scheme, these edge users suffer from grave
 PC, and hence their channel estimates have extremely poor quality, which
 severely limits the achievable UL detection performance. With the aid of the proposed SPR
 scheme, the full channel estimate at the BS of the $i$-th cell is then given by
\begin{equation}\label{eq24}
 \widehat{\mathbf{H}}_{i,i}=\big[\widehat{\mathbf{H}}_{i,i}^{\text{c}} ~
  \widehat{\mathbf{H}}_{i,i}^{\text{e}}\big] ,
\end{equation}
 which is significantly more accurate than that of the conventional
 channel estimation scheme of (\ref{eq6}). Thus, given this more
 accurate channel estimate, the UL achievable rate of the edge users can be
 significantly increased, which will be analyzed in detail in
 Section~\ref{S6}.

 Moreover, since the edge users of the adjacent cells rely on
 orthogonal pilot sequences, a BS can also acquire the partial
 inter-cell channels for the edge users of the adjacent
 cells. Specifically, by correlating the received pilot matrix
 $\overline{\mathbf{Y}}_i^{\text{p}}$ with
 $\mathbf{\Phi}_{\text{e},j}$, the BS of the $i$-th cell becomes capable of
 acquiring the partial inter-cell channels from the edge users in the
 $j$-th cell without PC, as follows:
\begin{equation}\label{eq25}
 \widehat{\mathbf{H}}_{i,j}^{\text{e}}=\frac{1}{\sqrt{\rho_{\text{p}}}}
 \overline{\mathbf{Y}}_i^{\text{p}}\mathbf{\Phi}_{\text{e},j}^H
 =\mathbf{H}_{i,j}^{\text{e}}+\overline{\mathbf{N}}_{i,j}^{\text{e}}, \hspace{0.2cm} j\neq i,
\end{equation}
 where
 $\overline{\mathbf{N}}_{i,j}^{\text{e}}=\frac{1}{\sqrt{\rho_{\text{p}}}}
 \overline{\mathbf{N}}_i^{\text{p}}\mathbf{\Phi}_{\text{e},j}^H$ can
 be rendered arbitrarily small upon increasing $M$. Thus, the BS of
 the $i$-th cell becomes capable of accurately estimating all the
 partial inter-cell channels of the links spanning from the edge users
 of the adjacent cells, which comprises an estimate of the inter-cell
 channel matrix
 $\mathbf{A}_i\in\mathbb{C}^{(K_{\text{e}}-K_{i,\text{e}})\times M}$
 as
\begin{equation}\label{eq26}
 \widehat{\mathbf{A}}_i = \big[\widehat{\mathbf{H}}_{i,1}^{\text{e}} \cdots
 \widehat{\mathbf{H}}_{i,i-1}^{\text{e}} ~ \widehat{\mathbf{H}}_{i,i+1}^{\text{e}}
 \cdots \widehat{\mathbf{H}}_{i,L}^{\text{e}}]^T .
\end{equation}
 For instance, in the simple example depicted in
 Fig.~\ref{InterCellChannel}, the BS in the 1st cell is able to
 acquire the accurate channel estimates of both its edge user as well
 as of the partial inter-cell channels of the edge users in two
 adjacent cells. The inter-cell channel matrix $\mathbf{A}_i$ provides
 important information for the DL transmit precoding design.
 Armed with its accurate estimate $\widehat{\mathbf{A}}_i$,
 the BS of the $i$-th cell will be able to beneficially preprocess its
 transmissions for the sake of reducing the ICI
 inflicted upon its neighbouring edge users roaming in the adjacent
 cells, which is the topic of the next subsection.

\subsection{Multi-Cell Block Diagonalization Precoding}\label{S5.3}

 By selecting the TPC vector for a specific user from
 the null space spanned by the channels of other users, the classical
 BD TPC scheme \cite{BD} adopted in single-cell multi-user
 MIMO systems is capable of eliminating the multi-user interference.
 Armed with the estimate of the partial inter-cell channels, we propose
 the MBD TPC by extending the classical BD TPC to
 a multi-cell multi-user scenario. Specifically, by projecting
 the DL transmit signal onto the null space of the subspace spanned
 by the inter-cell channels, the proposed MBD TPC becomes capable of
 eliminating the ICI imposed on these edge users.

 In order to obtain the null space of the inter-cell channels, we
 first apply the classic SVD \cite{SVD} to the inter-cell channel
 matrix $\widehat{\mathbf{A}}_i$, yielding
\begin{equation}\label{eq27}
 \widehat{\mathbf{A}}_i=\mathbf{U}_i\mathbf{\Sigma}_i\mathbf{V}_i^H,
\end{equation}
 where $\mathbf{U}_i\in\mathbb{C}^{(K_{\text{e}}-K_{i,\text{e}})\times(K_{\text{e}}-K_{i,\text{e}})}$
 denotes the left-singular-vector matrix, $\mathbf{V}_i\in\mathbb{C}^{M\times M}$ denotes
 the right-singular-vector matrix, and $\mathbf{\Sigma}_i\in
 \mathbb{C}^{(K_{\text{e}}-K_{i,\text{e}})\times M}$ is comprised of the singular values as
\begin{equation}\label{eq28}
 \mathbf{\Sigma}_i=\left[ \begin{array}{cc}
  \widehat{\mathbf{\Sigma}}_i & \mathbf{0}_{r_i\times(M-r_i)} \\
  \mathbf{0}_{(K_{\text{e}}-K_{i,\text{e}}-r_i)\times r_i} &
  \mathbf{0}_{(K_{\text{e}}-K_{i,\text{e}}-r_i)\times (M-r_i)}
 \end{array} \right],
\end{equation}
 in which $r_i=\text{rank}\big(\widehat{\mathbf{A}}_i\big)$ is the rank of $\widehat{\mathbf{A}}_i$,
 $\widehat{\mathbf{\Sigma}}_i=\text{diag}\big\{\sigma_{i,1},\sigma_{i,2}, \cdots ,
 \sigma_{i,r_i}\big\}$ and the singular values satisfy
\begin{equation}\label{eq29}
\sigma_{i,1}\geq \sigma_{i,2}\geq \cdots \geq \sigma_{i,r_i} > 0 .
\end{equation}
 According to the properties of full SVD, the null space of the inter-cell channels,
 namely, $\text{Null}\big(\widehat{\mathbf{A}}_i\big)\subseteq \mathbb{C}^{M}$, can
 be spanned by the columns of the matrix $\mathbf{B}_i\in\mathbb{C}^{M\times(M-r_i)}$,
 which is a sub-matrix of $\mathbf{V}_i$ defined by
\begin{equation}\label{eq30}
 \mathbf{B}_i=\big[\mathbf{v}_{i,r_i+1} ~ \mathbf{v}_{i,r_i+2} \cdots \mathbf{v}_{i,M}\big] ,
\end{equation}
 where $\mathbf{v}_{i,j}$ denotes the $j$-th column of
 $\mathbf{V}_i$. Note that the existence of this null space is
 guaranteed owing to the fact that the number of antennas at the BS of
 LS-MIMO systems is much larger than that of the edge users, i.e., we
 have
\begin{equation}\label{eq31}
 M\gg K_{\text{e}}\geq K_{\text{e}}-K_{i,\text{e}}\geq r_i.
\end{equation}
 The large null space of the inter-cell channels indicates that for any transmit precoding matrix
 chosen from this null space, i.e., for $\forall \mathbf{W}_i\subset\text{Null}(\widehat{\mathbf{A}}_i)$,
 we have
\begin{equation}\label{nullspace} 
 \widehat{\mathbf{A}}_i\mathbf{W}_i=\mathbf{0}\Rightarrow
 \big(\widehat{\mathbf{H}}_{i,j}^{\text{e}}\big)^T\mathbf{W}_i=\mathbf{0}, ~ \forall j\neq i,
\end{equation}
 which means that this transmit precoding matrix calculated for the
 $i$-th cell is capable of eliminating the ICI
 inflicted upon the edge users roaming in the adjacent cells.  It is
 plausible however that a precoding matrix, which is randomly chosen
 from the null space $\text{Null}\big(\widehat{\mathbf{A}}_i\big)$ may
 cause severe intra-cell interference.  To avoid the deleterious
 effects of intra-cell interference, we project a conventional
 transmit precoding matrix onto this null space.

 For example, by projecting this conventional MF precoding matrix $\mathbf{W}_i^{\text{MF}}=\frac{1}{\sqrt{\gamma_i^{\text{MF}}}}\widehat{\mathbf{H}}_{i,i}^{*}$ onto the null space
 $\text{Null}(\widehat{\mathbf{A}}_i)$, we can generate the MF based MBD matrix as
\begin{equation}\label{MFSMP} 
 \mathbf{W}_i^{\text{MFMBD}}=\frac{1}{\sqrt{\gamma_i^{\text{MFMBD}}}}
 \mathbf{P}_{\mathbf{B}_i}\widehat{\mathbf{H}}_{i,i}^{*} ,
\end{equation}
 where $\mathbf{P}_{\mathbf{B}_i}=\mathbf{B}_i\mathbf{B}_i^{\dagger}$ denotes the
 projection operator based on the matrix $\mathbf{B}_i$, and $\gamma_i^{\text{MFMBD}}$
 is a normalization factor given by
\begin{align}\label{eq35}
 \gamma_i^{\text{MFMBD}} =& \frac{1}{K_i}\text{Tr}\big(\widehat{\mathbf{H}}_{i,i}^T
 \mathbf{P}_{\mathbf{B}_i}^H\mathbf{P}_{\mathbf{B}_i}\widehat{\mathbf{H}}_{i,i}^*\big)
 = \frac{1}{K_i}\text{Tr}\big(\widehat{\mathbf{H}}_{i,i}^T\mathbf{P}_{\mathbf{B}_i}
 \widehat{\mathbf{H}}_{i,i}^*\big),
\end{align}
 in which $\mathbf{P}_{\mathbf{B}_i}^H=\mathbf{P}_{\mathbf{B}_i}$ and
 $\mathbf{P}_{\mathbf{B}_i}\mathbf{P}_{\mathbf{B}_i}=\mathbf{P}_{\mathbf{B}_i}$ are applied.

 Similarly, by projecting the conventional ZF transmit precoding matrix onto the null space
 $\text{Null}(\mathbf{A}_i)$, we can generate the ZF based MBD
 matrix as
\begin{align}\label{eq36}
 \mathbf{W}_i^{\text{ZFMBD}} =& \frac{1}{\sqrt{\gamma_i^{\text{ZFMBD}}}}\left(
 \big(\mathbf{P}_{\mathbf{B}_i}^*\widehat{\mathbf{H}}_{i,i}\big)^T\right)^{\dagger} \nonumber \\
 =& \frac{1}{\sqrt{\gamma_i^{\text{ZFMBD}}}}\mathbf{P}_{\mathbf{B}_i}
 \widehat{\mathbf{H}}_{i,i}^* \left(\widehat{\mathbf{H}}_{i,i}^T\mathbf{P}_{\mathbf{B}_i}^H
 \mathbf{P}_{\mathbf{B}_i}\widehat{\mathbf{H}}_{i,i}^*\right)^{-1} \nonumber \\
 =& \frac{1}{\sqrt{\gamma_i^{\text{ZFMBD}}}}\mathbf{P}_{\mathbf{B}_i}\widehat{\mathbf{H}}_{i,i}^*
 \left(\widehat{\mathbf{H}}_{i,i}^T\mathbf{P}_{\mathbf{B}_i}\widehat{\mathbf{H}}_{i,i}^*\right)^{-1} ,
\end{align}
 where the normalization factor $\gamma_i^{\text{ZFMBD}}$ is calculated as
\begin{align}\label{eq37}
 \gamma_i^{\text{ZFMBD}} =& \frac{1}{K_i}\text{Tr}\big(\widehat{\mathbf{H}}_{i,i}^T
 \mathbf{P}_{\mathbf{B}_i}^H\mathbf{P}_{\mathbf{B}_i}\widehat{\mathbf{H}}_{i,i}^*\big)^{-1} \nonumber \\
 =& \frac{1}{K_i}\text{Tr}\big(\widehat{\mathbf{H}}_{i,i}^T\mathbf{P}_{\mathbf{B}_i}
 \widehat{\mathbf{H}}_{i,i}^*\big)^{-1}.
\end{align}

\begin{figure}
\vspace*{1mm}
\center{\includegraphics[angle=0,width=0.4\textwidth]{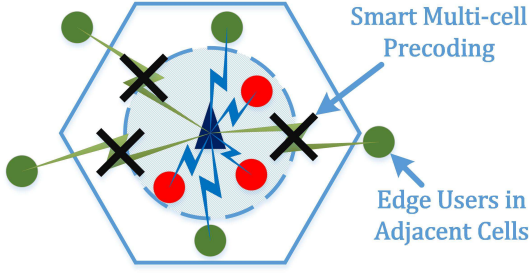}}
\caption{An illustrative example of the MBD precoding scheme.}
\label{NullSpacePrecoding} 
\vspace*{-3mm}
\end{figure}

 Both the precoding matrices $\mathbf{W}_i^{\text{MFMBD}}$ and
 $\mathbf{W}_i^{\text{ZFMBD}}$ are capable of eliminating the
 ICI imposed on the edge users of the adjacent
 cells. An illustrative example is depicted in
 Fig.~\ref{NullSpacePrecoding}, where based on the proposed SPR
 scheme, the BS becomes capable of estimating the partial inter-cell
 channels of the edge users roaming in the adjacent cells. The MBD TPC
 then projects the DL transmission signal onto the null space
 of the partial inter-cell channels in order to eliminate the
 ICI contaminating the reception of these edge
 users in the adjacent cells. Therefore, the MBD TPC
 significantly increases the DL achievable rate of edge users, and
 consequently the QoS of edge users is considerably enhanced.

\section{Performance Analysis}\label{S6}

 Before we investigate the performance of the proposed SPR and MBD
 schemes, we first discuss the amount of pilot resources required,
 then derive both the UL as well as the DL achievable rate, and finally
 consider the computational complexity imposed.

\subsection{Pilot Resource Consumption}\label{S6.1}

As seen in (\ref{KCS}), the conventional scheme requires
$K_{\text{CS}}=\max\{K_i,1\le i\le L\}$ number of orthogonal pilot
sequences and it suffers from grave PC. Again, in order to eliminate
the PC caused by the reuse of the same pilot group in adjacent cells,
the most plausible solution is to apply orthogonal pilot sequences to
all users in all cells. However, the number of orthogonal pilot
sequences would be increased to
\begin{equation}\label{eq38}
 K_{\text{OS}}=\sum_{i=1}^L K_i,
\end{equation}
 which leads to a substantial spectral efficiency reduction.

Recall that the proposed SPR and MBD schemes are capable of enhancing
the QoS for edge users at the expense of a slightly increased number
of pilot resources. More specifically, by comparing (\ref{KCS}) and
(\ref{KSPR}), the additional pilot resources required by the proposed
SPR scheme can be derived as
\begin{equation}\label{eq39}
 K_{\text{SPR}}-K_{\text{CS}}=\sum_{i\neq i_0}K_{i,\text{e}}\leq \sum_{i=1}^L K_{i,\text{e}},
\end{equation}
 where $i_0$ denotes the index of the cell which has the most users,
 i.e., $K_{i_0}=K_{\text{CS}}$.  It is clear that the additional number
 of pilot sequences is close to the total number of the edge users.
 Since the edge users are classified according to the threshold $\rho_i$,
 which can be adjusted by the parameter $\lambda$, the number of
 edge users can be flexibly adjusted too.

 More explicitly, the careful choice of the parameter $\lambda$
 provides a flexible trade-off between the pilot resources required
 and the achievable system performance of the proposed SPR scheme. At
 one extreme end, when the parameter $\lambda$ is set to 0, all the
 users will be regarded as edge users and the proposed SPR scheme
 becomes equivalent to the orthogonal scheme, where all users in all
 cells use orthogonal pilot sequences. Naturally, this achieves the
 best performance but relies on the most pilot resources, requiring
 $K_{\text{OS}}$ orthogonal pilot sequences. At the other extreme,
 when the parameter $\lambda$ is set to a sufficiently large value,
 e.g. $\lambda=K_{\text{CS}}$, then all users are regarded as center
 users and the proposed SPR scheme degrades to the conventional scheme
 that reuses the same pilot group in all cells. Hence the resultant
 arrangement attains the worst performance but consumes the minimum
 pilot resources, hence requiring only $K_{\text{CS}}$ orthogonal
 pilot sequences.

\subsection{Uplink Transmission}\label{S6.2}

 For the center users, the average SINR performance of SPR-aided UL
 transmission becomes almost the same as that of applying the
 conventional scheme. This is because for the center users in a cell,
 the estimated channel matrix of (\ref{eq6}) obtained by applying the
 conventional scheme is very similar to that of (\ref{eq22}) obtained
 by applying the SPR scheme. However, the achievable rate of
 the center users of the SPR-aided UL transmission is slightly
 reduced, since the pilot overhead increases, i.e.,
 $\mu\rightarrow\frac{K_{\text{SPR}}}{K_{\text{CS}}}\mu$.
 On the other hand, the performance of UL transmission for the
 edge users is much more complex, as seen below.

 Similar to the received signal given in section \ref{S3} based
 on the conventional scheme, the received signal at the BS of the
 $i$-th cell based on the SPR scheme can be represented as
\begin{equation}\label{eq40}
 \overline{\mathbf{y}}_i^{\text{u}} = \sqrt{\rho_{\text{u}}}\sum_{j=1}^L
 \big( \mathbf{H}_{i,j}^{\text{c}}\mathbf{x}_j^{\text{u,c}} + \mathbf{H}_{i,j}^{\text{e}}
 \mathbf{x}_j^{\text{u,e}} \big) + \overline{\mathbf{n}}_i^{\text{u}},
\end{equation}
 where $\mathbf{x}_j^{\text{u,c}}=\big[x_{j,1}^{\text{u,c}} ~ x_{j,2}^{\text{u,c}} \cdots
 x_{j,K_{j,\text{c}}}^{\text{u,c}}\big]^T$ denotes the symbol vector transmitted from the
 $K_{j,\text{c}}$ center users in the $j$-th cell, $\mathbf{x}_j^{\text{u,e}}=\big[x_{j,1}^{\text{u,e}}
 ~ x_{j,2}^{\text{u,e}} \cdots x_{j,K_{j,\text{e}}}^{\text{u,e}}\big]^T$ is the
 symbol vector transmitted from the $K_{j,\text{e}}$ edge users in the $j$-th cell, and
 $\overline{\mathbf{n}}_i^{\text{u}}$ denotes the corresponding UL AWGN vector.

 By adopting the MF detector based on the channel estimation obtained
 by the SPR scheme for the edge users in the $i$-th cell,
 i.e., $\widehat{\mathbf{H}}_{i,i}^{\text{e}}$ of (\ref{eq23}), the
 detected symbol vector for the $K_{i,\text{e}}$ users in the $i$-th
 cell is given by
\begin{align}\label{eq41}
 \widehat{\mathbf{x}}_i^{\text{u,e}} =& \big(\widehat{\mathbf{H}}_{i,i}^{\text{e}}\big)^H
 \overline{\mathbf{y}}_i^{\text{u}} \nonumber \\
 =& \big(\mathbf{H}_{i,i}^{\text{e}} +
 \overline{\mathbf{N}}_i^{\text{e}}\big)^H \left( \sqrt{\rho_{\text{u}}}
 \sum_{j=1}^L\big( \mathbf{H}_{i,j}^{\text{c}}\mathbf{x}_j^{\text{u,c}} +
 \mathbf{H}_{i,j}^{\text{e}}\mathbf{x}_j^{\text{u,e}}\big) + \overline{\mathbf{n}}_i^{\text{u}}
 \right) \nonumber\\
 =& \sqrt{\rho_{\text{u}}} \big(\mathbf{H}_{i,i}^{\text{e}}\big)^H \mathbf{H}_{i,i}^{\text{e}}
 \mathbf{x}_i^{\text{u,e}} + \boldsymbol{\eta}_i^{\text{u,e}}
 \overset{(a)}\approx M\sqrt{\rho_{\text{u}}}\mathbf{D}_{i,i}^{\text{e}}\mathbf{x}_i^{\text{u,e}} ,
\end{align}
 where $\mathbf{D}_{i,i}^{\text{e}}=\text{diag}\big\{\beta_{i,i,1}^{\text{e}},\beta_{i,i,2}^{\text{e}},
 \cdots , \beta_{i,i,K_{i,\text{e}}}^{\text{e}}\big\}$ denotes the sub-diagonal matrix of $\mathbf{D}_{i,i}$
 consisting of the $K_{i,\text{e}}$ edge users' large-scale fading coefficients, and
 $\boldsymbol{\eta}_i^{\text{u,e}}=\big[ \eta_{i,1}^{\text{u,e}} ~ \eta_{i,2}^{\text{u,e}} \cdots
 \eta_{i,K_{i,\text{e}}}^{\text{u,e}}\big]^T$ denotes the interference which can be made
 arbitrarily small by increasing the number of antennas at the BS. In particular, for the
 $k$-th edge user in the $i$-th cell, the detected symbol is given by
\begin{align}\label{eq42}
 \widehat{x}_{ik}^{\text{u,e}} =& \sqrt{\rho_{\text{u}}} \big(\mathbf{h}_{i,i,k}^{\text{e}}\big)^H
 \mathbf{h}_{i,i,k}^{\text{e}} x_{i,k}^{\text{u,e}} + \mu_{i,k}^{\text{u,e}} + \eta_{i,k}^{\text{u,e}} \nonumber \\
 \overset{(a)}\approx & M\sqrt{\rho_{\text{u}}} \beta_{i,i,k}^{\text{e}} x_{i,k}^{\text{u,e}},
\end{align}
 where $\mu_{i,k}^{\text{u,e}}=\sqrt{\rho_{\text{u}}} \sum_{k^{'}\neq
   k} \big( \mathbf{h}_{i,i,k}^{\text{e}}\big)^H
 \mathbf{h}_{i,i,k^{'}}^{\text{e}} x_{i,k^{'}}^{\text{u,e}}$ is the
 intra-cell interference arriving from the other edge users in the
 same cell, which can be rendered arbitrarily small by increasing
 $M$. Similar to the derivation in (\ref{SINRUL}), the UL SINR of the
 $k$-th edge user in the $i$-th cell can be calculated as
\begin{align}\label{UPSINRSPR} 
 \overline{\text{SINR}}_{i,k}^{\text{u,e}} =& \rho_{\text{u}}
 \frac{\big|\big(\mathbf{h}_{i,i,k}^{\text{e}}\big)^H\mathbf{h}_{i,i,k}^{\text{e}}\big|^2}
 {\big|\mu_{i,k}^{\text{u,e}}\big|^2+\big|\eta_{i,k}^{\text{u,e}}\big|^2} ,
\end{align}
 and the achievable UL rate can be calculated as
 $\overline{\text{C}}_{i,k}^{\text{u,e}}
 =(1-\frac{K_{\text{SPR}}}{K_{\text{CS}}}\mu)\text{E}\big\{\log_2\big(1+\overline{\text{SINR}}_{i,k}^{\text{u,e}}\big)\big\}$. Note that unlike the result of the conventional scheme given in
 (\ref{SINRUL}), $\overline{\text{SINR}}_{i,k}^{\text{u,e}}$ increases
 as $M$ increases, and in the asymptotic case of $M\rightarrow
 \infty$, we have $\overline{\text{SINR}}_{i,k}^{\text{u,e}}
 \rightarrow \infty$.

 In summary, in contrast to the conventional scheme which is unable to
 remove the PC by simply increasing the number of antennas at the BS
 $M$, for the edge users we eliminated the PC imposed on the UL data
 transmission and consequently the UL achievable rate is significantly
 improved. Similar results can be obtained if we adopt the ZF detector
 for UL transmission in our proposal, and the ZF detector outperforms
 the MF detector which will be verified by our numerical results.

\subsection{Downlink Transmission}\label{S6.3}

 Similar to the analysis of the UL transmission, in this section we
 focus our attention on the DL transmission of the edge users in our
 proposal.

 By adopting the MFMBD precoding matrix as derived in
 (\ref{MFSMP}), the received signal vector of the $K_{i,\text{e}}$
 edge users in the $i$-th cell can be represented as
\begin{align}\label{eq44}
 \overline{\mathbf{y}}_i^{\text{d,e}} =& \sqrt{\rho_{\text{d}}}\sum_{j=1}^L
 \big(\mathbf{H}_{j,i}^{\text{e}}\big)^T \mathbf{W}_j^{\text{MFMBD}}
 \left[ \begin{array}{c}
  \mathbf{x}_{j}^{\text{d,c}} \\
  \mathbf{x}_{j}^{\text{d,e}} \end{array} \right]
 + \overline{\mathbf{n}}_{i}^{\text{d,e}} \nonumber \\
 =& \sqrt{\rho_{\text{d}}} \sum_{j=1}^L \frac{1}{\sqrt{\gamma_j^{\text{MFMBD}}}}
 \big(\mathbf{H}_{j,i}^{\text{e}}\big)^T \mathbf{B}_j \mathbf{B}_j^{\dagger} \widehat{\mathbf{H}}_{j,j}^{*}
 \left[ \begin{array}{c}
  \mathbf{x}_{j}^{\text{d,c}} \\
  \mathbf{x}_{j}^{\text{d,e}} \end{array} \right]
 + \overline{\mathbf{n}}_{i}^{\text{d,e}} \nonumber \\
 \approx& \sqrt{ \frac{\rho_{\text{d}}} {\gamma_i^{\text{MFMBD}}} }
 \big(\mathbf{H}_{i,i}^{\text{e}}\big)^T \mathbf{B}_i \mathbf{B}_i^{\dagger} \widehat{\mathbf{H}}_{i,i}^{*}
 \left[ \begin{array}{c}
  \mathbf{x}_{i}^{\text{d,c}} \\
  \mathbf{x}_{i}^{\text{d,e}} \end{array} \right]
 + \overline{\mathbf{n}}_{i}^{\text{d,e}} ,
\end{align}
 where $\mathbf{x}_j^{\text{d,c}}=\big[x_{j,1}^{\text{d,c}} ~
   x_{j,2}^{\text{d,c}}
   \cdots,x_{j,K_{j,\text{c}}}^{\text{d,c}}\big]^T$ are the symbols
 transmitted to the $K_{j,\text{c}}$ center users in the $j$-th cell,
 $\mathbf{x}_j^{\text{d,e}}=\big[ x_{j,1}^{\text{d,e}} ~
   x_{j,2}^{\text{d,e}} \cdots
   x_{j,K_{j,\text{e}}}^{\text{d,e}}\big]^T$ are the symbols destined for the
 $K_{j,\text{e}}$ edge users in the $j$-th cell,
 $\overline{\mathbf{n}}_{i}^{\text{d,e}}
 =\big[\overline{n}_{i,1}^{\text{d,e}} ~
   \overline{n}_{i,2}^{\text{d,e}} \cdots
   \overline{n}_{i,K_{i,\text{e}}}^{\text{d,e}}\big]^T$ denotes the
 corresponding DL AWGN vector, and the approximation $\approx$ holds
 as we apply
 $\widehat{\mathbf{H}}_{j,i}^{\text{e}}\approx\mathbf{H}_{j,i}^{\text{e}}$
 and
 $\big(\widehat{\mathbf{H}}_{j,i}^{\text{e}}\big)^T\mathbf{B}_j=\mathbf{0}$
 for $j\neq i$ (see (\ref{nullspace})).

 Thus, for the $k$-th edge user in the $i$-th cell, the received symbol can be represented as
\begin{align}\label{eq45}
 \overline{y}_{i,k}^{\text{d,e}} \approx& \sqrt{\frac{\rho_{\text{d}}}{\gamma_i^{\text{MFMBD}}}}
 \big(\mathbf{h}_{i,i,k}^{\text{e}}\big)^T \mathbf{B}_i \mathbf{B}_i^{\dagger}
 \big(\mathbf{h}_{i,i,k}^{\text{e}}\big)^* x_{i,k}^{\text{d,e}} + \mu_{i,k}^{\text{d,e}}
 + \overline{n}_{i,k}^{\text{d,e}},
\end{align}
 where the intra-cell interference $\mu_{i,k}^{\text{d,e}}$ is given in (\ref{eq46}).
\begin{figure*}[!tp]
\begin{align}\label{eq46}
 \mu_{i,k}^{\text{d,e}} = &\sqrt{\frac{\rho_{\text{d}}} {\gamma_i^{\text{MFMBD}}}}
 \left( \sum_{k^{'}=1}^{K_{i,\text{c}}} \big(\mathbf{h}_{i,i,k}^{\text{e}}\big)^T
 \mathbf{B}_i \mathbf{B}_i^{\dagger} \big(\mathbf{h}_{i,i,k^{'}}^{\text{c}}\big)^*
 x_{i,k^{'}}^{\text{d,c}} + \sum_{k^{'}\neq k} \big(\mathbf{h}_{i,i,k}^{\text{e}}\big)^T
 \mathbf{B}_i \mathbf{B}_i^{\dagger} \big(\mathbf{h}_{i,i,k^{'}}^{\text{e}}\big)^*
 x_{i,k^{'}}^{\text{d,e}}\right) .
\end{align}
\end{figure*}
 Both $\overline{n}_{i,k}^{\text{d,e}}$ and $\mu_{i,k}^{\text{d,e}}$ can be made
 arbitrarily small by increasing the number of antennas at the BS. The DL SINR of the
 $k$-th edge user in the $i$-th cell can then be calculated as
\begin{equation}\label{eq47}
 \overline{\text{SINR}}_{i,k}^{\text{d,e}} \approx \frac{\rho_{\text{d}}} {\gamma_i^{\text{MFMBD}}}
 \cdot \frac{\big|\big(\mathbf{h}_{i,i,k}^{\text{e}}\big)^T \mathbf{B}_i \mathbf{B}_i^{\dagger}
 \big(\mathbf{h}_{i,i,k}^{\text{e}}\big)^*\big|^2}{\big|\mu_{i,k}^{\text{d,e}}\big|^2
 +\big|\overline{n}_{i,k}^{\text{d,e}}\big|^2},
\end{equation}
 and the achievable DL rate can be calculated as
 $\overline{\text{C}}_{i,k}^{\text{d,e}}=(1-\frac{K_{\text{SPR}}}{K_{\text{CS}}}\mu)\text{E}\big\{\log_2\big(1+\overline{\text{SINR}}_{i,k}^{\text{d,e}}\big)\big\}$.
 In contrast to the result of the conventional scheme given in
 (\ref{SINRDL}), $\overline{\text{SINR}}_{i,k}^{\text{d,e}}$ increases
 as $M$ increases, and in the asymptotic case of $M\rightarrow
 \infty$, hence we have $\overline{\text{SINR}}_{i,k}^{\text{d,e}}
 \rightarrow \infty$.

 It is clear that for the edge users the PC is eliminated by our
 proposed scheme during the DL data transmission, and additionally,
 both the ICI as well as intra-cell interference
 imposed on these edge users has been reduced by the MBD scheme.
 Similar results can be obtained if we adopt the ZF based MBD
 TPC matrix, i.e., $\mathbf{W}_i^{\text{ZFMBD}}$, for DL
 transmission in our proposal.

 Taking into account the extra pilot resource, the achievable
 DL rate of the edge users has been significantly improved, while
 that of the center users is slightly reduced.  Moreover, the
 average UL and DL cell throughput of the SPR and MBD assisted
 system will be confirmed later by our simulation study.

\subsection{Computational Complexity}\label{S6.4}

 The computational complexity of implementing the MBD scheme
 at the BS for the edge users will be quantified in terms of the
 number of complex-valued multiplications required, which includes the
 following two main contributions:
\begin{enumerate}
\item For the SVD operator, the complexity is on the order of $M
  K_{\text{e}}^2$, which is denoted by $\mathcal{O}\big(M
  K_{\text{e}}^2\big)$, which allows us to calculate
  $\widehat{\mathbf{A}}_i=
  \mathbf{U}_i\mathbf{\Sigma}_i\mathbf{V}_i^H$ by using the QR
  decomposition.
\item For the matrix pseudo inverse operation, the complexity is on
  the order of $\mathcal{O}\big(M K_{\text{CS}}^2\big)$, which alloes
  us to generate the ZF based MBD precoding matrix by using the Gram-Schmidt
  algorithm.
\end{enumerate}

 The total computational complexity of implementing the MBD
 scheme at the BS is therefore on the order of $\mathcal{O}\big(M\big(K_{\text{e}}^2+K_{\text{CS}}^2\big)\big)$,
 which is comparable to that of the conventional scheme and it is within the computational
 capability of a typical state-of-the-art BS.

\begin{table}
\vspace*{-1mm}
\centering
\caption{Basic Simulation Parameters}
\label{parameters} 
\vspace*{-1mm}
\begin{tabular}{l|l}
  \hline\hline
  Number of cells $L_{\text{total}}$ & 19 \\ \hline
  Number of antennas in BS $M$ & $32\leq M\leq 256$ \\ \hline
  Number of users in the $i$-th cell $K_i$ & $8\leq K_i\leq 10$ \\ \hline
  Number of pilot resource $K_{\text{CS}}$ & $K_{\text{CS}}=10$ \\ \hline
  Threshold $\rho_i$ adjustment parameter $\lambda$ & $0.05\leq\lambda\leq 1$ \\ \hline
  Cell radius $R$ & 500 m \\ \hline
  Average transmit power at users $\rho_{\text{p}},\rho_{\text{u}}$ & 10 dBm \\ \hline
  Average transmit power at BS $\rho_{\text{d}}$ & 12 dBm \\\hline
  Path loss exponent $\alpha$ & 3 \\ \hline
  Log normal shadowing fading $\sigma_{\text{shadow}}$ & 8 dB \\ \hline
  Carrier frequency & 2 GHz \\ \hline
  System bandwidth & 10 MHz \\ \hline
  Thermal noise density & -174 dBm/Hz\\ \hline
  Pilot overhead parameter $\mu$ & $\mu=0.1$ \\ \hline
  Minimum distance between user and BS & 30 m \\ \hline
  \hline
\end{tabular}
\end{table}

\section{Simulation Study}\label{S7}

\begin{figure}
\vspace*{-3mm}
\center{\includegraphics[width=0.48\textwidth]{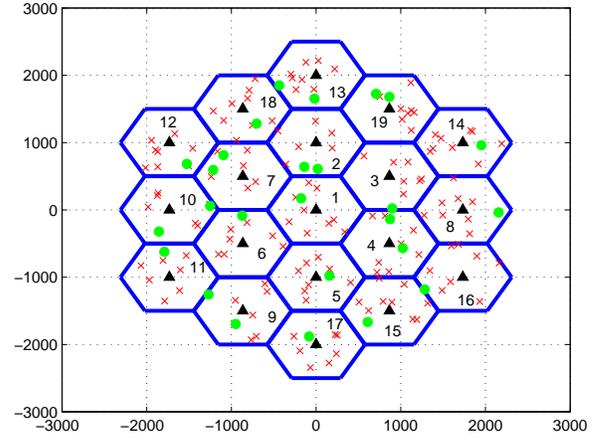}}
\vspace*{-4mm}
\caption{An instantiation of the randomly generated user distribution
  in the simulated hexagonal cellular network, where red crosses,
  green dots, and black numbers denote center users, edge users, and cell numbers, respectively.}
\label{UserPosition} 
\vspace*{-3mm}
\end{figure}

 We evaluated the performance of the proposed SPR and MBD schemes
 using a set of Monte-Carlo simulations. A typical hexagonal
 cellular network of $L_{\text{total}}$ cells was considered, where the BS of each
 cell employed $M$ AEs and the $i$-th cell had $K_i$
 single-AE users \cite{scalingupMIMO, cell191, cell192}.
 The default values of the various parameters of
 this simulated hexagonal cellular network are summarized in
 Table~\ref{parameters}.
 The large-scale fading coefficient $\beta_{i,j,k}$ was generated
 according to \cite{scalingupMIMO}
\begin{equation}\label{eq48}
 \beta_{i,j,k}=\frac{z_{i,j,k}}{(r_{i,j,k}/R)^{\alpha}},
\end{equation}
 where $R$ denotes the cell radius, and $\alpha$ is the path loss
 exponent, while $r_{i,j,k}$ is the distance between the $k$-th user
 in the $j$-th cell and the BS in the $i$-th cell, while $z_{i,j,k}$
 denotes the shadow fading factor, which obeys the log-normal
 distribution, i.e., $10\log_{10}\big(z_{i,j,k}\big)$ follows the
 zero-mean Gaussian distribution having a standard deviation of
 $\sigma_{\text{shadow}}$.
 The reuse factor of center pilot group $\bm{\Phi}_{\text{c}}$ is 1,
 i.e., it is reused in all $L_{\text{total}}$ cells, while the reuse factor of edge
 sub pilot groups $\bm{\Phi}_{\text{e}}=\left[\mathbf{\Phi}_{\text{e},1}^T ~ \mathbf{\Phi}_{\text{e},2}^T
 \cdots \mathbf{\Phi}_{\text{e},7}^T\right]^T$ is 7, i.e., the $i$-th cell utilizes $\mathbf{\Phi}_{\text{e},\text{mod}(i,7)+1}$ and non adjacent cells
 reuse the same edge sub pilot group. The locations of the users in each cell
 were all randomly generated in each trial. A particular simulation
 trial is shown in Fig.~\ref{UserPosition}, where the red crosses and
 green dots in each cell denote the center users and
 edge users, respectively, which are classified by
 the BS based on the threshold $\rho_i$ associated with the parameter $\lambda=0.1$.
 As mentioned previously and also seen from Fig.~\ref{UserPosition}, the
 classification of center users and edge users is not based on their
 distance from the serving BS.

\begin{figure}
\vspace*{-1mm}
\center{\includegraphics[width=0.48\textwidth]{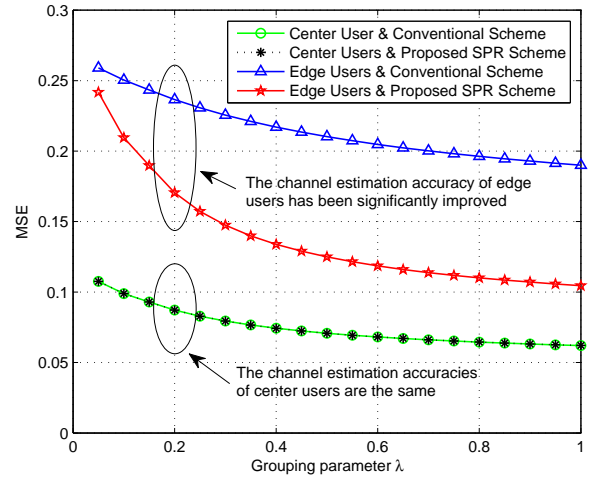}}
\vspace*{-4mm}
\caption{Channel estimation accuracy comparison for the conventional and proposed SPR schemes with $M=128$.}
\label{CE_MSE_Lambda} 
\vspace*{-3mm}
\end{figure}

 Fig.~\ref{CE_MSE_Lambda} compares the channel estimation accuracies as
 functions of the grouping parameter $\lambda$ for both the conventional and
 the proposed SPR schemes with $M=128$. In each simulation trial, the channel
 estimation mean square error (MSE) of the edge users is calculated as
\begin{equation}\label{eq49}
 \text{MSE}^{\text{e}}=\text{E}\bigg\{\frac{1}{K_{\text{e}}}\sum_{i=1}^{L}\sum_{k=1}^{K_{i,\text{e}}}
 \frac{\big\|\widehat{\mathbf{h}}_{i,i,k}^{\text{e}}-\mathbf{h}_{i,i,k}^{\text{e}}\big\|^2_2}
 {\big\|\mathbf{h}_{i,i,k}^{\text{e}}\big\|^2_2}\bigg\},
\end{equation}
 where $\widehat{\mathbf{h}}_{i,i,k}^{\text{e}}$ denotes the estimate of the true
 channel vector $\mathbf{h}_{i,i,k}^{\text{e}}$, while the MSE of the channel estimation for
 the center users is defined as
\begin{equation}\label{eq50}
 \text{MSE}^{\text{c}}=\text{E}\bigg\{\frac{1}{\big(\sum_{i=1}^{L}K_i\big)-K_{\text{e}}}
 \sum_{i=1}^{L}\sum_{k=1}^{K_{i,\text{c}}}
 \frac{\big\|\widehat{\mathbf{h}}_{i,i,k}^{\text{c}}-\mathbf{h}_{i,i,k}^{\text{c}}\big\|^2_2}
 {\big\|\mathbf{h}_{i,i,k}^{\text{c}}\big\|^2_2}\bigg\},
\end{equation}
 where $\widehat{\mathbf{h}}_{i,i,k}^{\text{c}}$ denotes the estimate
 of the true channel vector $\mathbf{h}_{i,i,k}^{\text{c}}$. The
 average results over 100 random simulation runs are presented in
 Fig.~\ref{CE_MSE_Lambda}. By increasing the grouping parameter $\lambda$,
 more users will be regarded as edge users.
 As expected, for the center users, who only
 suffer from a slight PC, both the conventional and proposed SPR
 schemes attain the same excellent channel estimation
 accuracy. However, the conventional scheme attains a poor channel
 estimation accuracy for the edge users, who suffer from severe pilot
 contamination. By contrast, since the PC is eliminated by applying
 orthogonal pilot sequences for the edge users in the adjacent cells,
 the channel estimation accuracy achieved by the proposed SPR scheme
 is significantly improved.

\begin{figure}
\vspace*{-3mm}
\center{\includegraphics[width=0.48\textwidth]{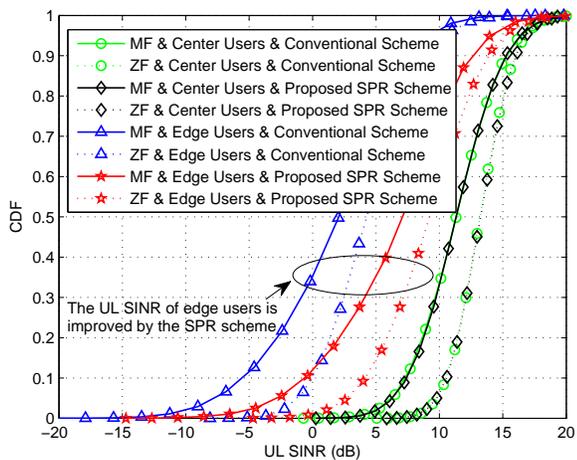}}
\vspace*{-4mm}
\caption{The CDF of UL SINR for the conventional and proposed SPR
 schemes with $M=128$ and $\lambda=0.1$.}
\label{SINR_UL_CDF} 
\vspace*{-3mm}
\end{figure}

 Fig.~\ref{SINR_UL_CDF} shows the cumulative density function (CDF) of UL SINR for both the
 conventional and for the proposed SPR schemes with $M=128$ and $\lambda=0.1$,
 where the results are presented by 1000 random simulation trials. In each simulation run,
 the conventional scheme calculates the UL SINR of a center or an edge
 user according to the first line of (\ref{ULdatadetection}). For our SPR scheme, the UL SINR of
 center users is similar to that for the conventional scheme, while the
 UL SINR of edge users is calculated by the first line of (\ref{eq41}) in each
 simulation trial. Since the UL transmission of the proposed SPR scheme is almost
 the same as that of the conventional scheme for the center users,
 their curves in Fig.~\ref{SINR_UL_CDF} are almost coincided.
 Observed in Fig.~\ref{SINR_UL_CDF}, our SPR scheme attains a significantly
 higher UL SINR for the edge users than the conventional scheme.
 Furthermore, ZF detector is always better than MF detector by about 2 dB
 for both center users and edge users.

\begin{figure}
\vspace*{-3mm}
\center{\includegraphics[width=0.48\textwidth]{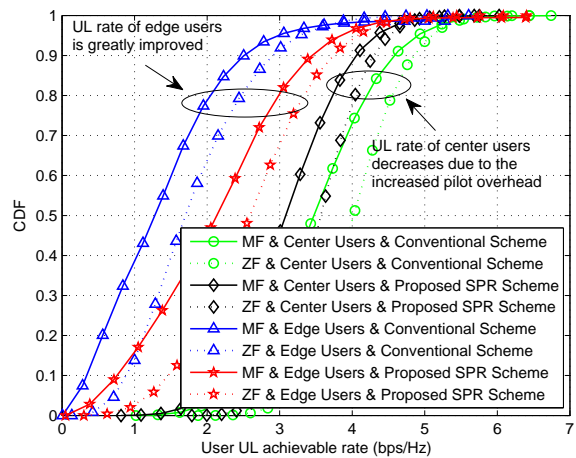}}
\vspace*{-4mm}
\caption{The CDF of UL achievable rate for the conventional and proposed SPR
 schemes with $M=128$ and $\lambda=0.1$.}
\label{UL_C_CDF} 
\vspace*{-3mm}
\end{figure}

 Fig.~\ref{UL_C_CDF} shows the CDF of UL achievable rate for the
 conventional and proposed SPR schemes with $M=128$ and $\lambda=0.1$,
 where the results are obtained from 500 random simulation runs.
 In each simulation trial, we generate the user positions first and then generate the channels of users
 for 50 times to obtain the UL achievable rate.
 Although the UL SINR results of the center users of the conventional and SPR
 schemes are the same as shown in Fig.~\ref{SINR_UL_CDF}, the UL achievable rates
 are different due to the different pilot overhead, i.e.,
 $\mu\rightarrow\frac{K_{\text{SPR}}}{K_{\text{CS}}}\mu$.
 It is clear that the UL achievable rate of edge users is significantly improved
 by the SPR scheme, while the UL achievable rate of center users decreases due
 to the increased pilot overhead.
 Moreover, the ZF detector always outperforms the MF detector by about 0.3 bps/Hz per user.

\begin{figure}
\center{\includegraphics[width=0.48\textwidth]{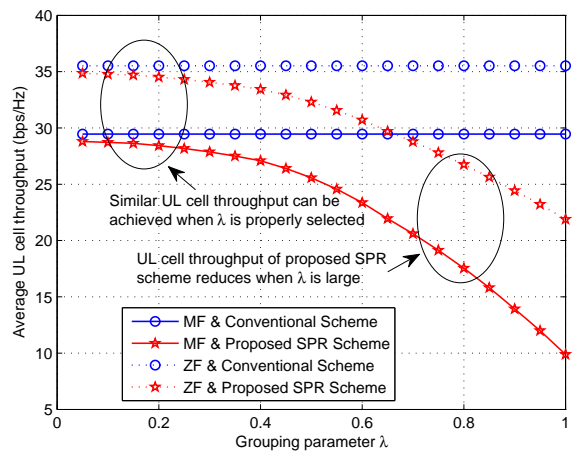}}
\vspace*{-4mm}
\caption{The average UL cell throughput for the conventional and proposed SPR
 schemes with $M=128$ against $\lambda$.}
\label{UL_C_lambda} 
\vspace*{-3mm}
\end{figure}

 Fig.~\ref{UL_C_lambda} shows the average UL cell throughput for the
 conventional and proposed SPR schemes with $M=128$. It is clear that
 by increasing the group parameter $\lambda$, there will be more
 users regarded as edge users, which leads to the increase of pilot overhead
 and the decrease of UL achievable rate of center users.
 Thus, the proper selection of grouping parameter is important, e.g., $\lambda\leq 0.2$,
 which both improves the performance of edge users and also ensures the cell throughput.
 Otherwise, e.g., $\lambda>0.5$, it is clear that the loss caused by
 over large pilot overhead outweights the gain of the SPR scheme.
 In addition, the ZF detector always provides a gain about 5 bps/Hz of
 average cell throughput compared with the MF detector.

\begin{figure}
\vspace*{-4mm}
\center{\includegraphics[width=0.48\textwidth]{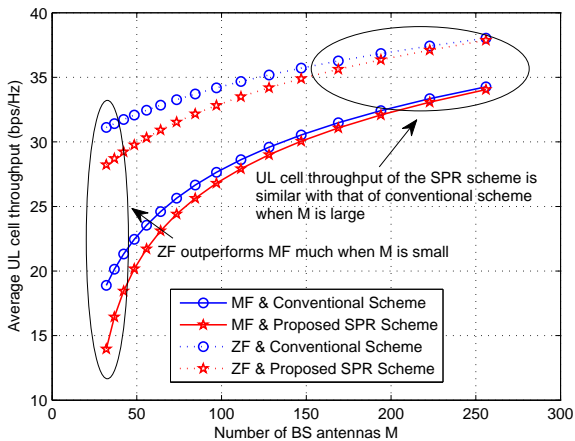}}
\vspace*{-4mm}
\caption{The average UL cell throughput for the conventional and proposed SPR
 schemes with $\lambda=0.1$ against $M$.}
\label{UL_C_M} 
\vspace*{-3mm}
\end{figure}

 Fig.~\ref{UL_C_M} shows the average UL cell throughput comparison of
 the conventional and proposed SPR schemes with $\lambda=0.1$ against
 the number of BS antennas $M$. When the number of BS antennas is small,
 i.e., $M=32$, the average UL cell throughput of the proposed SPR scheme
 is smaller than that of the conventional scheme about 5 bps/Hz with
 MF detector adopted. It is obvious that, to obtain the performance
 gain for edge users as shown in Fig.~\ref{UL_C_CDF}, the proposed SPR
 scheme scarifies the spectral efficiency due to the increased pilot
 overhead and leads to the average UL cell throughput reduction.
 However, by increasing the number of BS antennas, e.g., $M=256$, it
 becomes clear that the average UL cell throughput of the proposed
 SPR scheme approaches that of the conventional scheme since the
 performance of edge users can be significantly improved by increasing $M$.
 Moreover, the ZF detector outperforms the MF detector a lot when $M$
 is small, and the gap shrinks as $M$ increasing.

\begin{figure*}
\vspace*{-4mm}
\center{\includegraphics[width=1\textwidth]{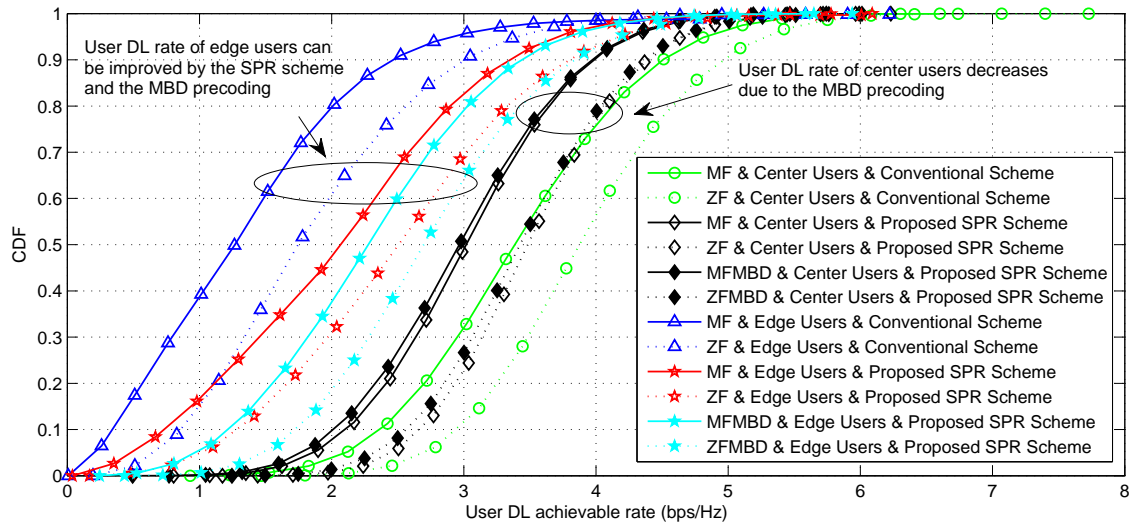}}
\vspace*{-4mm}
\caption{The CDF of DL achievable rate for the conventional system,
the SPR aided system as well as the SPR and MBD assisted system
with $M=128$ and $\lambda=0.1$.}
\label{DL_C_CDF} 
\vspace*{-3mm}
\end{figure*}

 Fig.~\ref{DL_C_CDF} shows the CDF of DL achievable rate for the
 conventional system, the SPR aided system as well as the SPR and MBD
 assisted system with $M=128$ and $\lambda=0.1$. Despite the MBD
 precoding scheme, it is clear that the results of DL achievable rate
 are similar with that of UL achievable rate as shown in
 Fig.~\ref{UL_C_CDF} due to their duality property.  When the MBD
 precoding is considered, we find that the DL achievable rate of edge
 users can be significantly improved due to the elimination of the
 ICI, while the DL achievable rate of center users slightly decreases
 since the projecting operator of the MBD precoding sacrifices degrees
 of freedom of the DL signals for center users. Again, we can find
 that the ZFMBD precoding achieves a gain about 0.4 bps/Hz compared
 with the MFMBD precoding for edge users.

\begin{figure}
\vspace*{-4mm}
\center{\includegraphics[width=0.48\textwidth]{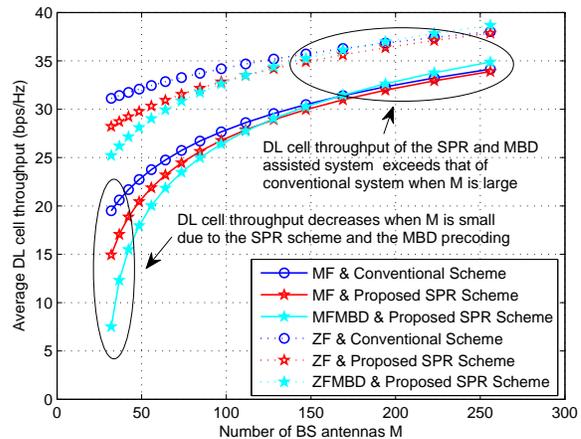}}
\vspace*{-4mm}
\caption{The average DL cell throughput for the conventional system, the SPR
aided system as well as the SPR and MBD assisted system with $\lambda=0.1$ against $M$.}
\label{DL_C_M} 
\vspace*{-3mm}
\end{figure}

 Fig.~\ref{DL_C_M} shows the average DL cell throughput for the conventional
 system, the SPR aided system as well as the SPR and MBD assisted system
 with $\lambda=0.1$ against $M$. The conventional system outperforms the
 SPR aided system, while the SPR and MBD assisted system performs worst
 when small number of BS antennas is considered, e.g., $M=32$.
 However, considering the typical massive MIMO configuration as $M=256$,
 it is clear that the average DL cell throughput of the SPR aided system
 approaches that of the conventional scheme, and the SPR and MBD assisted
 system performs best of all three. Moreover, when $M$ is further increased,
 the performance gap between the SPR and MBD assisted system and the conventional
 system will also become larger, which means that the increased rate of
 edge users becomes larger than the decreased rate of center users.

\section{Conclusions}\label{S8}

 We have developed a soft pilot reuse and multi-cell block diagonalization
 precoding regime for LS-MIMO systems, which are capable of
 significantly enhancing both the achievable UL and DL rate
 for edge users. Our contribution is twofold. Firstly, we break away from the
 traditional practice of treating all users as though they suffer from
 the same level of PC, and propose a simple yet effective means of
 dividing the users into cell-center and cell-edge users. This
 grouping allows us to apply the proposed SPR scheme, whereby a center
 pilot group is reused for the center users in all cells, while the
 edge pilot group is applied to the edge users in the adjacent cells.
 By requiring a slightly increased number of pilot sequences, the
 proposed SPR scheme eliminates the pilot contamination inflicted upon
 the edge users who would otherwise suffer from severe PC in the
 conventional scheme. This significantly enhances the QoS for the edge
 users, meanwhile ensures both the average UL and DL cell throughput
 with slight and negligible reduction compared with that of the conventional system.
 Secondly, we further exploit the fact that the BS becomes capable of
 estimating the inter-cell channels of the edge users in the adjacent
 cells with the aid of the our SPR regime without the deliterious
 effects of PC. Finally, we extend the classical BD precoding to a
 multi-cell scenario and propose the MBD precoding to eliminate the
 ICI imposed on the edge users of the adjacent cells in the DL. This
 MBD precoding further enhanced the performance of edge users in DL
 transmission and improved the average DL cell throughput in addition
 to the gain obtained by the SPR scheme.

\begin{IEEEbiography}[{\includegraphics[width=0.9in,height=1.1in,clip,keepaspectratio]{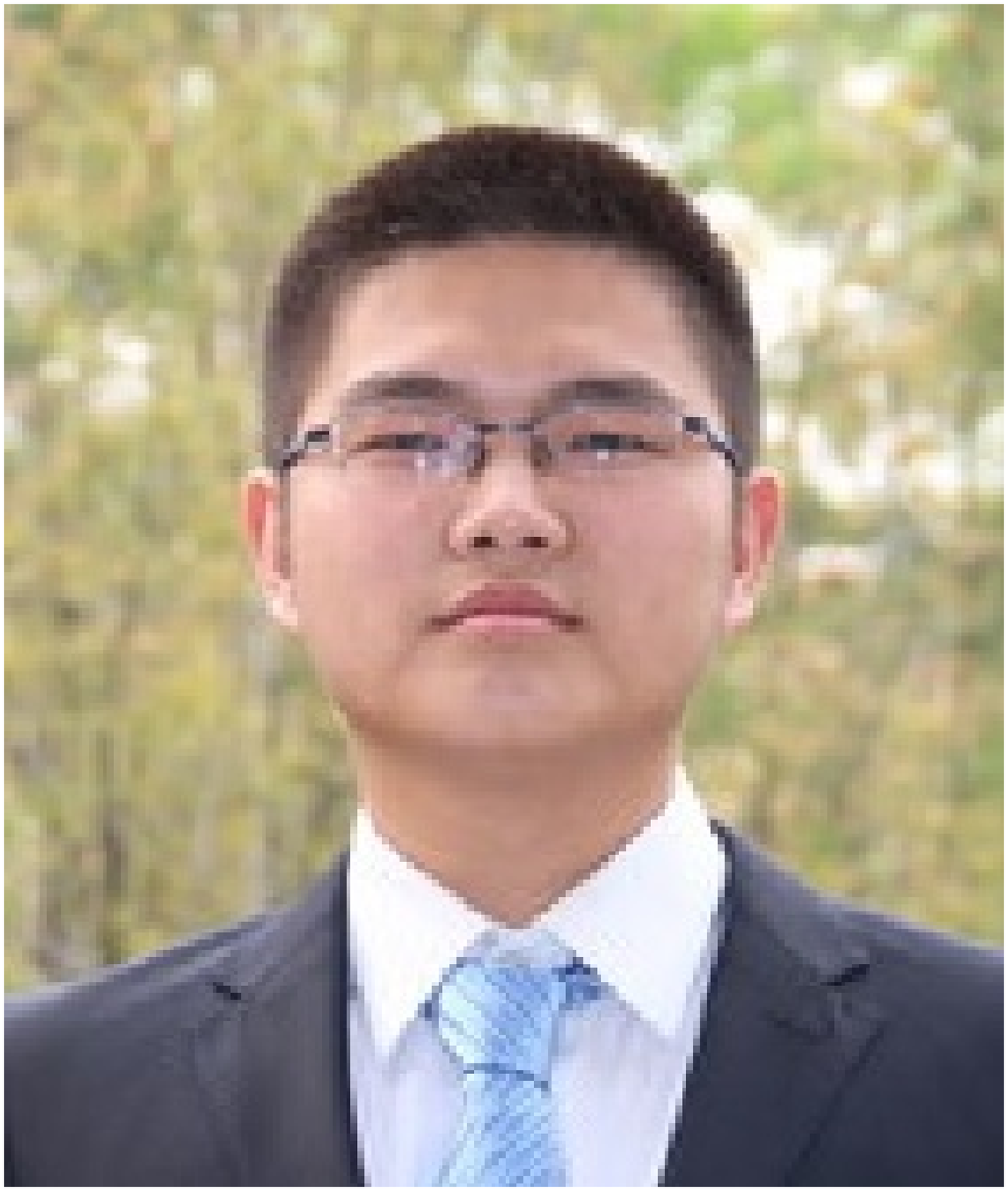}}]{Xudong Zhu}
received his B.S. degree from the department of Electronic Engineering in Tsinghua University, Beijing, China, in 2012.
He has been pursuing the Ph.D. degree at the Broadband Communication and Signal Processing Laboratory in Tsinghua University from 2012 to now.

His main research interests are in the areas of MIMO technique and mobile communication, especially in sparse signal reconstruction, massive MIMO, channel estimation, and precoding technique.
\end{IEEEbiography}

\begin{IEEEbiography}[{\includegraphics[width=0.9in,height=1.1in,clip,keepaspectratio]{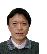}}]{Zhaocheng Wang}
(M'09-SM'11) received his B.S., M.S. and Ph.D. degrees from Tsinghua University in 1991, 1993 and 1996, respectively. From 1996 to 1997, he was with Nanyang Technological University (NTU) in Singapore as a Post Doctoral Fellow. From 1997 to 1999, he was with OKI Techno Centre (Singapore) Pte. Ltd., firstly as a research engineer and then as a senior engineer. From 1999 to 2009, he worked at SONY Deutschland GmbH, firstly as a senior engineer and then as a principal engineer.

He is currently a Professor at the Department of Electronic Engineering, Tsinghua University. His research areas include wireless communications, visible light communications, millimeter wave communications and digital broadcasting. He holds 34 granted US/EU patents and has published over 80 SCI indexed journal papers. He has served as technical program committee co-chairs of many international conferences. He is a Senior Member of IEEE and a Fellow of IET.
\end{IEEEbiography}

\begin{IEEEbiography}[{\includegraphics[width=0.9in,height=1.1in,clip,keepaspectratio]{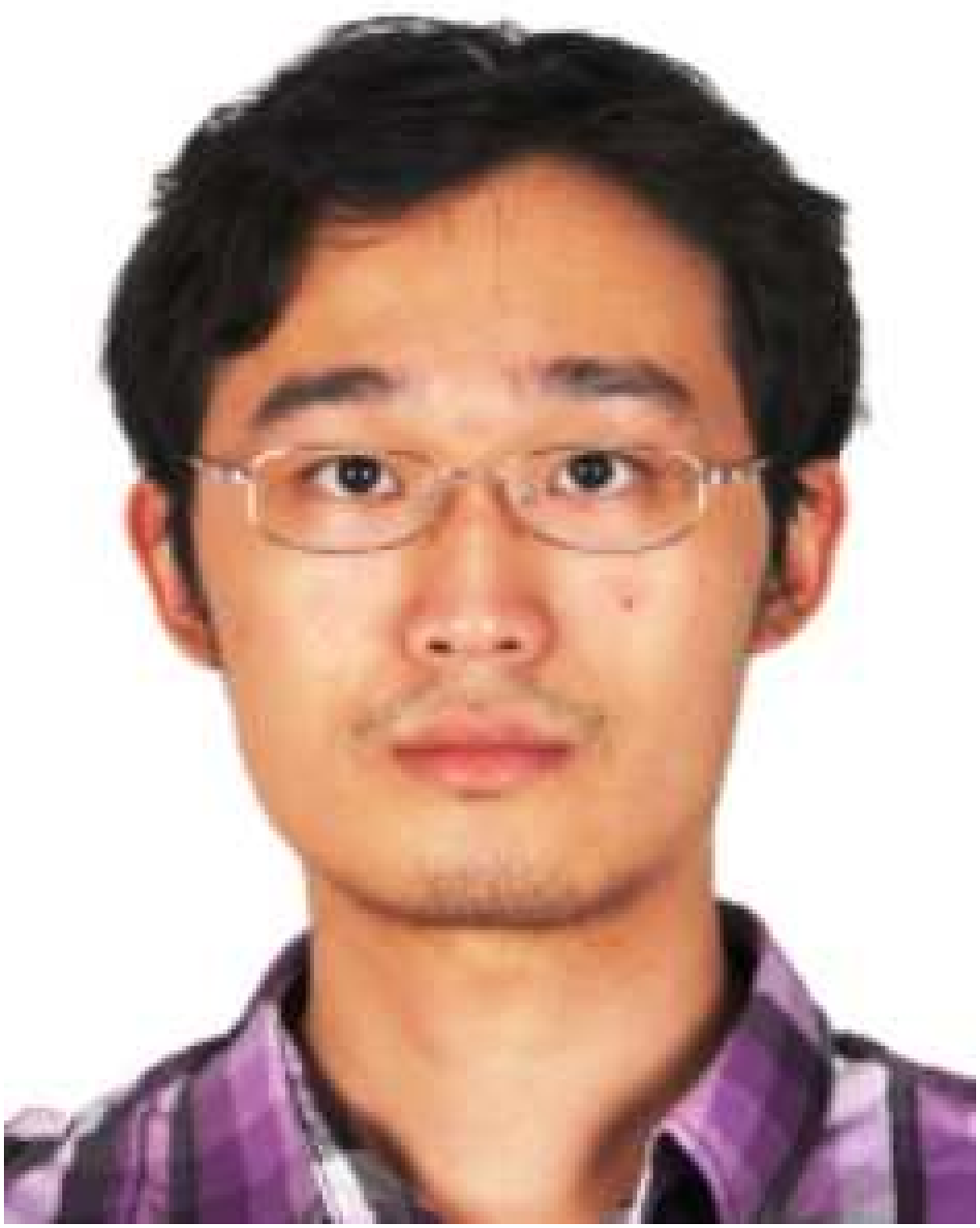}}]{Chen Qian}
received his B.S. degree from the department of Electronic Engineering in Tsinghua University, Beijing, China, in 2010.
He is now a PhD candidate of the Broadband Communication and Signal Processing Laboratory in Tsinghua University from 2010.
His research area includes channel coding technology, MIMO detection technology, and massive MIMO technology.
\end{IEEEbiography}

\begin{IEEEbiography}[{\includegraphics[width=0.9in,height=1.1in,keepaspectratio]{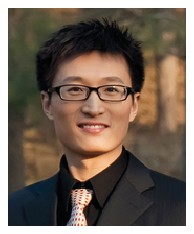}}]{Linglong Dai}
(M'11--SM'14) received the B.S. degree from Zhejiang University in 2003, the M.S. degree (with the highest honor) from China Academy of Telecommunications Technology (CATT) in 2006, and the Ph.D. degree (with the highest honors) from Tsinghua University, Beijing, China, in 2011.

From 2011 to 2013, he was a Postdoctoral Fellow at the Department of Electronic Engineering, Tsinghua University, and then since July 2013, became an Assistant Professor with the same Department. His research interests are in wireless communications with the emphasis on OFDM, MIMO, synchronization, channel estimation, multiple access techniques, and wireless positioning.

He has published over 50 journal and conference papers. He has received IEEE Scott Helt Memorial Award in 2015 (IEEE Transactions on Broadcasting Best Paper Award), IEEE ICC Best Paper Award in 2014, URSI Young Scientists Award in 2014, National Excellent Doctoral Dissertation Nomination Award in 2013, IEEE ICC Best Paper Award in 2013, Excellent Doctoral Dissertation of Beijing in 2012, Outstanding Ph.D. Graduate of Tsinghua University in 2011.
\end{IEEEbiography}

\begin{IEEEbiography}[{\includegraphics[width=0.9in,height=1.1in,keepaspectratio]{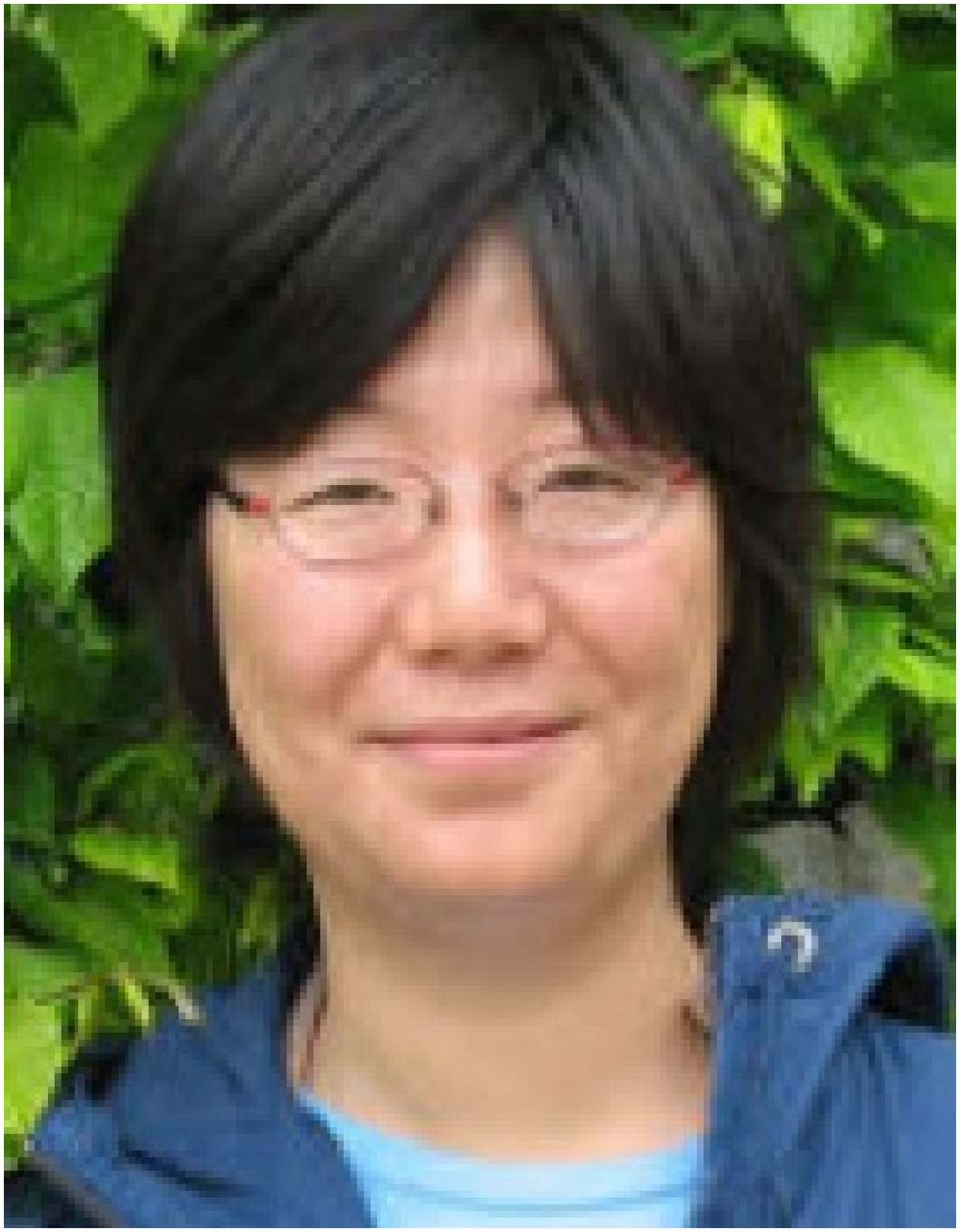}}]{Jinhui Chen}
is a researcher and Assistant Manager with Sony China Research Laboratory, Sony Corporation, China.

She obtained the Ph.D. degree from Telecom ParisTech, France, in 2009. From 2006 to 2009, she had been Research Assistant of Prof. Dirk T. M. Slock at EURECOM, France. From 2009 to 2014, she had been Research Scientist with Alcatel-Lucent Bell Labs China. She joined Sony in 2014. Her current research interests include: MIMO systems and wireless networks.
\end{IEEEbiography}

\begin{IEEEbiography}[{\includegraphics[width=0.9in,height=1.1in,clip,keepaspectratio]{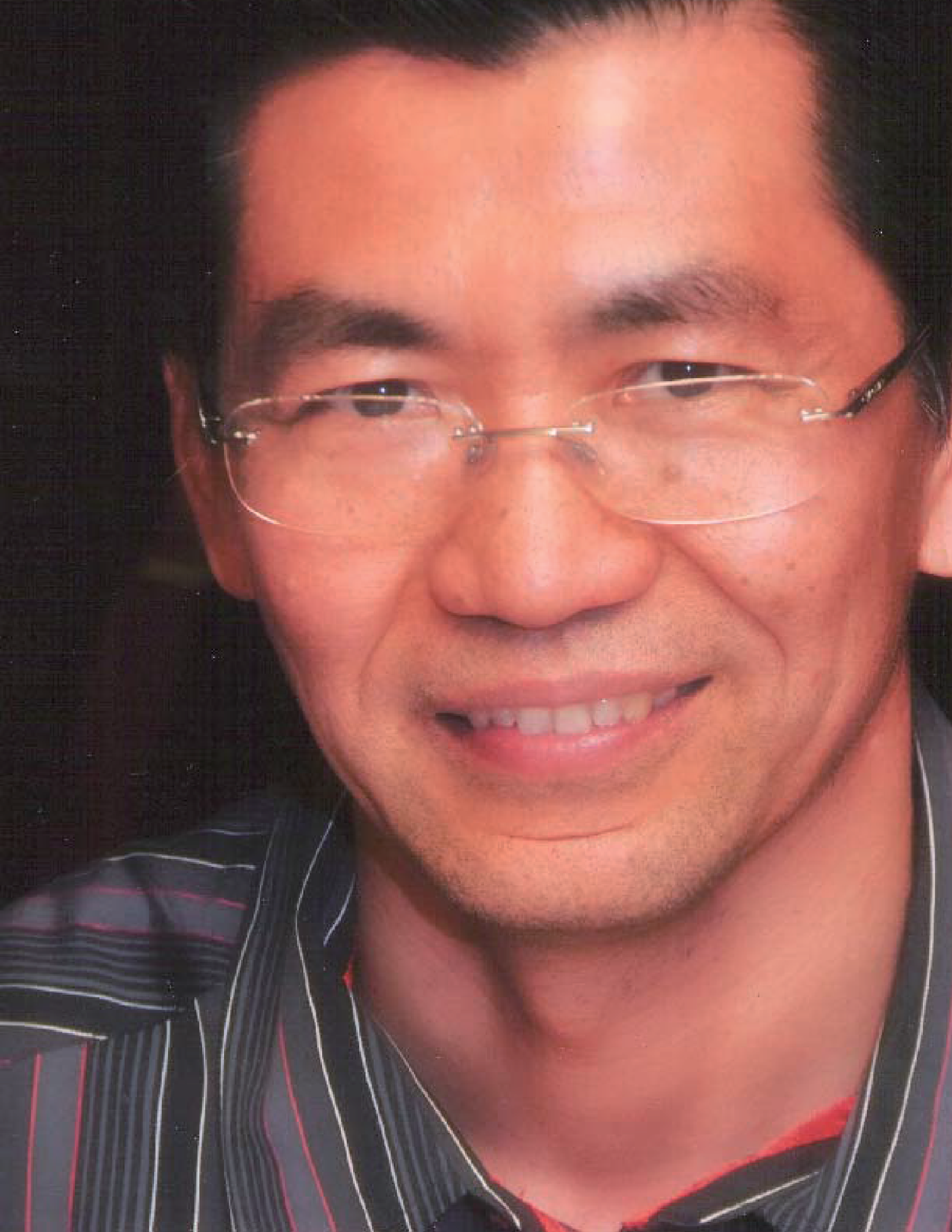}}]{Sheng Chen}
(M'90-SM'97-F'08) received his BEng degree from the East China Petroleum Institute, Dongying, China, in 1982, and his PhD degree from the City University, London, in 1986, both in control engineering. In 2005, he was awarded the higher doctoral degree, Doctor of Sciences (DSc), from the University of Southampton, Southampton, UK.

From 1986 to 1999, He held research and academic appointments at the Universities of Sheffield, Edinburgh and Portsmouth, all in UK. Since 1999, he has been with Electronics and Computer Science, the University of Southampton, UK, where he currently holds the post of Professor in Intelligent Systems and Signal Processing. Dr Chen's research interests include adaptive signal processing, wireless communications, modelling and identification of nonlinear systems, neural network and machine learning, intelligent control system design, evolutionary computation methods and optimisation. He has published over 500 research papers.

Dr. Chen is a Fellow of IET, a Distinguished Adjunct Professor at King Abdulaziz University, Jeddah, Saudi Arabia, and an ISI highly cited researcher in engineering (March 2004). He was elected to a Fellow of the United Kingdom Royal Academy of Engineering in 2014.
\end{IEEEbiography}

\begin{IEEEbiography}[{\includegraphics[width=0.9in,height=1.1in,clip,keepaspectratio]{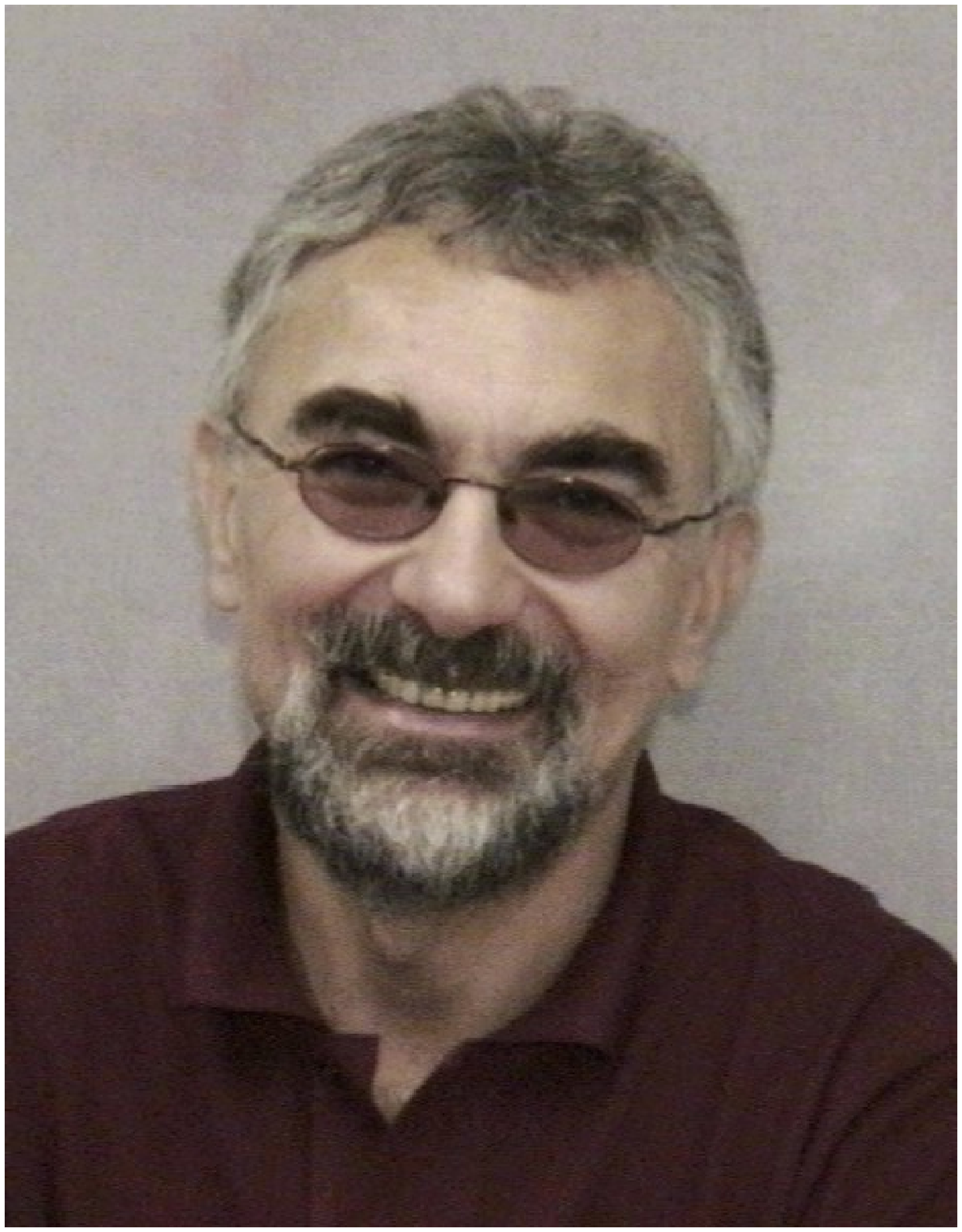}}]{Lajos Hanzo}
(http://www-mobile.ecs.soton.ac.uk) FREng, FIEEE, FIET, Fellow of EURASIP, DSc received his degree in electronics in 1976 and his doctorate in 1983. In 2009 he was awarded the honorary doctorate ``Doctor Honoris Causa'' by the Technical University of Budapest.

During his 38-year career in telecommunications he has held various research and academic posts in Hungary, Germany and the UK. Since 1986 he has been with the School of Electronics and Computer Science, University of Southampton, UK, where he holds the chair in telecommunications.  He has successfully supervised about 100 PhD students, co-authored 20 John Wiley/IEEE Press books on mobile radio communications totalling in excess of 10 000 pages, published 1400+ research entries at IEEE Xplore, acted both as TPC and General Chair of IEEE conferences, presented keynote lectures and has been awarded a number of distinctions.

Currently he is directing a 100-strong academic research team, working on a range of research projects in the field of wireless multimedia communications sponsored by industry, the Engineering and Physical Sciences Research Council (EPSRC) UK, the European Research Council's Advanced Fellow Grant and the Royal Society's Wolfson Research Merit Award.  He is an enthusiastic supporter of industrial and academic liaison and he offers a range of industrial courses.

He is also a Governor of the IEEE VTS.  During 2008 - 2012 he was the Editor-in-Chief of the IEEE Press and a Chaired Professor also at Tsinghua University, Beijing.  His research is funded by the European Research Council's Senior Research Fellow Grant.  For further information on research in progress and associated publications please refer to http://www-mobile.ecs.soton.ac.uk Lajos has 20 000+ citations.
\end{IEEEbiography}

\begin{thebibliography}{1}

\bibitem{noncooperative} 
 T.~L.~Marzetta, ``Noncooperative cellular wireless with unlimited numbers of base station
 antennas,'' \textit{IEEE Trans. Wirel. Commun.}, vol.~9, no.~11, pp.~3590--3600, Nov. 2010.

\bibitem{scalingupMIMO} 
 F.~Rusek, D.~Persson, B.~K.~Lau, E.~G.~Larsson, T.~L.~Marzetta, O.~Edfors, and F.~Tufvesson,
 ``Scaling up MIMO: Opportunities and challenges with very large arrays,'' \textit{IEEE
 Signal Process. Mag.}, vol.~30, no.~1, pp.~40--60, Jan 2013.

\bibitem{LuLu} 
 L.~Lu, G.~Y.~Li, A.~L.~Swindlehurst, A.~Ashikhmin, and R.~Zhang, ``An overview of massive MIMO:
 Benefits and challenges,'' \textit{IEEE J. Sel. Top. Signal Process.}, vol.~8, no.~5,
 pp.~742--758, Oct. 2014.

\bibitem{milliwatts} 
 H.~Q.~Ngo, E.~G.~Larsson, and T.~L.~Marzetta, ``Energy and spectral efficiency of very large
 multiuser MIMO systems,'' \textit{IEEE Trans. Commun.}, vol.~61, pp.~1436--1449, Apr. 2013.

\bibitem{conventionalMIMO} 
 A.~Goldsmith, S.~A.~Jafar, N.~Jindal, and S.~Vishwanath, ``Capacity limits of MIMO channels,''
 \textit{IEEE J. Sel. Areas Commun.}, vol.~21, no.~5, pp.~684--702, Jun. 2003.

\bibitem{TDD1} 
 J.~Jose, A.~Ashikhmin, T..~L.~Marzetta, and S.~Vishwanath, ``Pilot contamination and
 precoding in multi-cell TDD systems,'' \textit{IEEE Trans. Wirel. Commun.}, vol.~10, no.~8,
 pp.~2640--2651, Aug. 2011.

\bibitem{FFR} 
 F. Jin, R. Zhang, and L. Hanzo, ``Frequency-swapping aided femtocells in twin-layer cellular networks relying on fractional frequency reuse," \textit{IEEE Wirel. Commun. and Network. Conf. (WCNC), 2012}, April 1-4, 2012, pp.~3097--3101.

\bibitem{2G5G} 
 Y. Zhou, L. Liu, H. Du, L. Tian, X. Wang, and J. Shi, ``An overview on intercell interference management in mobile cellular networks: From 2G to 5G," \textit{IEEE Int. Conf. Commun. Systems (ICCS), 2014}, 19-21 Nov., pp.~217--221, 2014.

\bibitem{ComP1} 
L.~Daewon, S.~Hanbyul, B.~Clerckx, E.~Hardouin, D.~Mazzarese, S.~Nagata, and K.~Sayana, ``Coordinated multipoint transmission and reception in LTE-advanced: deployment scenarios and operational challenges," \textit{IEEE Commun. Mag.}, vol.~50, no.~2, pp.~148--155, Feb. 2012.

\bibitem{ComP2} 
C.~Junil, Z.~Chance, D.~J.~Love, and U.~Madhow, ``Noncoherent trellis coded quantization: A practical limited feedback technique for massive MIMO systems," \textit{IEEE Trans. Commun.}, vol.~61, no.~12, pp.~5016--5029, Dec. 2013.

\bibitem{timeshift1} 
 K.~Appaiah, A.~Ashikhmin, and T.~L.~Marzetta, ``Pilot contamination reduction in multi-user TDD systems,'' in \textit{Proc. ICC 2010} (Cap Town, South Africa), May 23-27, 2010, pp.~1--5.

\bibitem{timeshift2} 
 F.~Fernandes, A.~Ashikhmin, and T.~L.~Marzetta, ``Inter-cell interference in noncooperative
 TDD large scale antenna systems,'' \textit{IEEE J. Sel. Areas Commun.}, vol.~31, no.~2,
 pp.~192--201, Feb. 2013.

\bibitem{PCP1} 
 H.~Huh, S.-H.~Moon, Y.-T.~Kim, I.~Lee, and G.~Caire, ``Multi-cell MIMO downlink with cell
 cooperation and fair scheduling: A large-system limit analysis,'' \textit{IEEE Trans. Inf.
 Theory}, vol.~57, no.~12, pp.~7771--7786, Dec. 2011.

\bibitem{PCP2} 
 A.~Ashikhmin and T.~Marzetta, ``Pilot contamination precoding in multi-cell large scale
 antenna systems,'' in \textit{Proc. 2012 IEEE Int. Symp. Information Theory} (Cambridge, MA),
 July 1-6, 2012, pp.~1137--1141.

\bibitem{channelcovariance} 
 H.~Q.~Ngo and E.~G.~Larsson, ``EVD-based channel estimation in multicell multiuser MIMO
 systems with very large antenna arrays,'' in \textit{Proc. ICASSP 2012}( Kyoto, Japan),
 March 25-30, 2012, pp.~3249--3252.

\bibitem{AOA1} 
 H.~Yin, D.~Gesbert, M.~Filippou, and Y.~Liu, ``A coordinated approach to channel
 estimation in large-scale multiple-antenna systems,'' \textit{IEEE J. Sel. Areas Commun.},
 vol.~31, no.~2, pp.~264--273, Feb. 2013.

\bibitem{AOA2} 
 H.~Yin, D.~Gesbert, M.~Filippou, and Y.~Liu, ``Decontaminating pilots in massive MIMO
 systems,'' in \textit{Proc. ICC 2013} (Budapest, Hungary), June 9-13, 2013, pp.~3170--3175.

\bibitem{dataaided} 
 J.~Ma and P.~Li, ``Data-aided channel estimation in large antenna systems,'' \textit{IEEE
 Trans. on Signal Process.}, vol.~62, no.~12, pp.~3111--3124, Jun. 2014.

\bibitem{blind1} 
 L.~Cottatellucci, R.~R.~Muller, and M.~Vehkapera, ``Analysis of pilot decontamination based
 on power control,'' in \textit{Proc. VTC Spring 2013} (Dresden, Germany), June 2-5, 2013, pp.~1--5.

\bibitem{blind2} 
 R.~R.~Muller, L.~Cottatellucci, and M.~Vehkapera, ``Blind pilot decontamination,'' \textit{IEEE J.
 Sel. Top. Signal Process.}, vol.~8, no.~5, pp.~773--786, Oct. 2014.

\bibitem{Zhang_etal2014} 
 J.~Zhang, B.~Zhang, S.~Chen, X.~Mu, M.~El-Hajjar, and L.~Hanzo, ``Pilot contamination elimination
 for large-scale multiple-antenna aided OFDM systems,'' \textit{IEEE J. Sel. Top. Signal Process.},
 vol.~8, no.~5, pp.~759--772, Oct. 2014.

\bibitem{SVD} 
 R.~Persico, \textit{The Singular Value Decomposition}. Wiley-IEEE Press, 2014.

\bibitem{BD} 
 Q.~H.~Spencer, A.~L.~Swindlehurst, and M.~Haardt, ``Zero-forcing methods for downlink spatial multiplexing in multiuser MIMO channels," \textit{IEEE Trans. Signal Process.}, vol.~52, no.~2, pp.~461--471, Feb. 2004.

\bibitem{Overhead} 
 B.~Hassibi and B.~M.~Hochwald, ``How much training is needed in multiple-antenna wireless links?," \textit{IEEE Trans. Inf. Theory}, vol.~49, no.~4, pp.~951--963, Apr. 2003.

\bibitem{cell191} 
 Y.~P.~Seung, C.~Junil, and D.~J.~Love, ``Multicell cooperative scheduling for two-tier cellular networks," \textit{IEEE Trans. Commun.}, vol.~62, no.~2, pp.~536--551, Feb. 2014.

\bibitem{cell192} 
 Y.~J.~Hong, N.~Lee, and B. Clerckx, ``System level performance evaluation of inter-cell interference coordination schemes for heterogeneous networks in LTE-A system," \textit{IEEE GLOBECOM Workshops (GC Wkshps), 2010}, Dec. 6-10, pp.~690--694, 2010.
\end{thebibliography}
\end{document}